\documentclass[smallextended]{svjour3}       
\smartqed  
\usepackage{graphicx}%
 \usepackage{mathptmx}      

\usepackage{bm}
\usepackage{amssymb}
\usepackage{amsmath}
\usepackage{natbib}
\bibpunct[, ]{(}{)}{;}{a}{}{,}

\newcommand{\aB}{a_\mathrm{B}}
\newcommand{\aion}{a_\mathrm{i}}
\newcommand{\alphaf}{\alpha_\mathrm{f}}
\newcommand{\am}{a_\mathrm{m}}
\newcommand{\chiA}{\chi^\mathrm{A}}
\newcommand{\ChiA}{\bm{\chi}^\mathrm{A}}
\newcommand{\chiH}{\chi^\mathrm{H}}
\newcommand{\ChiH}{\bm{\chi}^\mathrm{H}}
\newcommand{\dd}{\mathrm{d}}
\newcommand{\EF}{\epsilon_\mathrm{F}}
\newcommand{\gcc}{\mbox{g~cm$^{-3}$}}
\newcommand{\gfact}{g_\mathrm{i}}
\newcommand{\Kc}{K_\mathrm{c}}
\newcommand{\khat}{\hat{\mathbf{k}}}
\newcommand{\lambdi}{\lambda_\mathrm{i}}
\newcommand{\mel}{m_\mathrm{e}}
\newcommand{\mion}{m_\mathrm{i}}
\newcommand{\Mspin}{(2\sion+1)}
\newcommand{\nion}{n_\mathrm{i}}
\newcommand{\omc}{\omega_\mathrm{c}}
\newcommand{\omci}{\omega_\mathrm{ci}}
\newcommand{\ompe}{\omega_{\mathrm{pe}}}
\newcommand{\opac}{\varkappa}
\newcommand{\pF}{p_\mathrm{F}}
\newcommand{\req}[1]{Eq.~(\ref{#1})}
\newcommand{\rhos}{\rho_\mathrm{s}}

\newcommand{\sion}{s_\mathrm{i}}
\newcommand{\sSB}{\sigma_\mathrm{SB}}

\newcommand{\Tc}{T_\mathrm{crit}}
\newcommand{\Teff}{T_\mathrm{eff}}

\newcommand{\Ts}{T_\mathrm{s}}
\newcommand{\xg}{x_\mathrm{g}}
\newcommand{\ycol}{y_\mathrm{col}}
\newcommand{\zete}{\zeta_\mathrm{e}}
\newcommand{\zeti}{\zeta_\mathrm{i}}
\newcommand{\Znuc}{Z_\mathrm{n}}

\journalname{Space Science Reviews}
\begin{document}

\title{Neutron Stars -- Thermal Emitters\thanks{The final
publication is available at Springer via
http://dx.doi.org/10.1007/s11214-014-0102-2}
}

\author{Alexander~Y.~Potekhin \and
Andrea De Luca         \and
Jos\'e~A.~Pons
}

\authorrunning{A.Y. Potekhin, A. De Luca, J.A. Pons} 

\institute{
   A.Y.~Potekhin \at
      Ioffe Institute,
      Politekhnicheskaya 26, 194021 Saint Petersburg,
       Russia;\\
      Central Astronomical Observatory at Pulkovo,
      Pulkovskoe Shosse 65, 196140 Saint Petersburg, Russia\\
              \email{palex@astro.ioffe.ru}
\and
A.~De Luca \at
  INAF -- Istituto di Astrofisica Spaziale e Fisica Cosmica
Milano,
     via E. Bassini 15, 20133, Milano, Italy;\\
  INFN -- Istituto Nazionale di Fisica Nucleare, sezione di
Pavia,
    via A.~Bassi 6, 27100, Pavia, Italy\\
    \email{deluca@iasf-milano.inaf.it}
\and
    J.A.~Pons \at
     Departament de F\'{\i}sica Aplicada, Universitat d'Alacant,
    Ap. Correus 99, E-03080 Alacant, Spain\\
      \email{jose.pons@ua.es}
}

\date{Received: 3 July 2014 / Accepted: 11 September 2014}

\maketitle

\begin{abstract}
Confronting theoretical models with observations of thermal
radiation emitted by neutron stars is one of the most
important ways to understand the properties of both, 
superdense matter in the interiors of the neutron stars and
dense magnetized plasmas in their outer layers.
Here we review the theory of thermal emission from the
surface layers of  strongly magnetized neutron stars, and
the main properties of the observational data. In
particular, we focus on the nearby sources for which a clear
thermal component has been detected, without being
contaminated by other emission processes (magnetosphere,
accretion, nebulae). We also discuss the applications of the
modern theoretical models of the formation of spectra of
strongly magnetized neutron stars to the observed thermally
emitting objects.
\end{abstract}

\section{Introduction}
\label{intro}

One of the first expectations of neutron-star (NS) astrophysics, dating back to the epoch when
such sources where theoretically proposed, is the possibility to detect thermal 
radiation from their hot (millions of K) surfaces \citep{Zwicky38}. 
Observation of the first 
sources of cosmic X-rays in the Early Sixties suggested that
NSs could indeed be detected, which added 
impetus to theoretical work as well as to modeling of expected NS observational properties.
The crucial point, attracting  a lot of interest in the astrophysical community, 
was that the study of NS cooling could constrain the physical properties of superdense matter 
in the interiors of the NSs \citep{Bahcall65}, 
under conditions that cannot be studied in terrestrial laboratories. 
Discovery of NSs as radio pulsars \citep{Hewish68} 
further boosted NS cooling studies as well as observational
efforts. However, the first detection of thermal radiation from the surface of INSs 
\citep{Cheng83,Brinkmann87} had to wait 
the launch of focusing X-ray telescopes
HEAO2/\textit{Einstein}
(1978--1981) and \textit{EXOSAT} (1983--1986), which
allowed for a dramatic leap forward in sensitivity to faint point X-ray sources.

Our understanding of the physics of NSs dramatically improved in the last 
decades. Thanks to multiwavelength observations (with an important role played 
by  X-ray observations), we have discovered with much surprise that not all Isolated NSs (INSs)
behave as radio pulsars, but that there exist a rich diversity of INS classes,
including Rotation-Powered Pulsars, Rotating Radio
Transients, Anomalous X-ray Pulsars, Soft Gamma Repeaters,
Central Compact Objects in supernova remnants, and the Magnificent Seven thermally
emitting NSs (see the next section).
It is now commonly accepted that magnetic fields plays an important role in shaping
the electromagnetic emission properties of INSs, e.g. mediating the conversion
of their rotational energy into radiation (as in rotation-powered sources), or directly
acting as the energy reservoir for most of the INS luminosity (as in magnetars).

One of the main challenges is to disentangle different emission components, overlapping
in the X-ray energy range. After discriminating thermal emission, a detailed study of the 
thermal spectra can yield precious information about the NS surface temperature
distribution, the properties of dense magnetized plasmas in their envelopes
and atmospheres, as well as set constraints on the equation of state of the
ultradense matter in the NS cores. 

Besides INSs, some NSs with observed thermal spectra also reside in binary systems.
In low-mass X-ray binary systems (LMXBs), a NS
accretes matter from a less massive star (a Main Sequence
star or a white dwarf), alternating periods of
intense accretion and periods of quiescence. When accretion
stops and the residual heat diffuses out from the crust, X-ray radiation comes from the
heated NS surface \citep{BBR98}. During the last decade, such
quiescent sources (qLMXBs) yield ever increasing amount of
valuable information on the NSs. 
Their spectra are successfully interpreted 
with models of NS atmospheres 
\citep[see][for a discussion and references]{P14}.
Another class of NSs in binaries with thermal spectra 
are X-ray bursters -- accreting  NSs in close binary systems, which
produce X-ray bursts with intervals from hours to days
\citep[see, e.g.][for a review]{StrohmayerBildsten}.
During intervals between the bursts, a burster's atmosphere
does not essentially differ from an atmosphere of a cooling
NS. In such periods, the bulk of the observed
X-ray radiation arises from transformation of gravitational
energy of the accreting matter into thermal energy.
Some of the bursts (so-called long bursts, which last over a
minute) occur during periods when the accretion rate is low enough for the luminosity
not to exceed a few percent of the Eddington limit. In this cases, a thermal atmospheric 
spectrum can be observed \citep[e.g.,][]{Suleimanov-ea11}.

As we will see, the number of known NSs with an
unambiguously identified thermal component in the spectrum
is not large, but it steadily increases.  Some of them can
be understood with models of nonmagnetic atmospheres,
whereas others are believed to be endowed with strong
magnetic fields,  which must be taken into account. After
the seminal work of \citet{Romani87}, the nonmagnetic
neutron-star atmospheres have been studied in many works
(see \citealp{Zavlin09} for a review). Databases of
neutron-star hydrogen atmosphere model spectra have been
published
\citep{ZavlinPS96,GaensickeBR02,Heinke-ea06},\footnote{Models
\textsc{NSA}, \textsc{NSAGRAV}, and \textsc{NSATMOS} in the
database \textit{XSPEC} \citep{XSPEC}.} and a computer code
for their calculation has been released
\citep{Haakonsen-ea12}. A database of carbon atmosphere
model spectra has been also published recently
\citep{Suleimanov-ea14}.\footnote{Model \textsc{CARBATM} in
the database \textit{XSPEC} \citep{XSPEC}.}  Model spectra
were calculated for neutron-star atmospheres composed of
different chemical elements from H to Fe
\citep[e.g.,][]{RajagopalRomani96,Pons-ea02,Heinke-ea06,HoHeinke09}
and mixtures of different elements
\citep{GaensickeBR02,Pons-ea02}. In general, the thermal
spectra of NSs in binaries (e.g., bursters and qLMXBs
mentioned above) are interpreted with theoretical models
without magnetic fields. Thermal components of spectra of
several millisecond pulsars (e.g., PSR J0437$-$4715,
\citealp{Bogdanov13} and some thermally emitting INSs (e.g.,
the source in Cassiopeia A, \citealp{HoHeinke09}) have also
been interpreted with the nonmagnetic atmosphere models. In
this paper, however, we will not consider the nonmagnetic
models (see \citealt{P14} for a more general review, which
includes a discussion of both the nonmagnetic and magnetic
cases), but instead we will focus on models of thermal
spectra of the INSs that are significantly affected by
strong magnetic fields.

The paper is organized as follows. In
Sect.~\ref{sect:INSfamilies} we give account of the
different classes of INSs. In Sect.~\ref{sec:objects} we
list INSs with confirmed thermal emission and give a summary
of their main characteristics. In
Sect.~\ref{sect:field-effects} we consider the definitions
and concepts that are important for the theory of formation
of thermal spectra of NSs with strong magnetic fields. The
latter theory is described in Sect.~\ref{sect:atm}.
Section~\ref{sect:cond} is devoted to the theory of a
magnetized condensed surface as an alternative to gaseous NS
atmospheres. In Sect.~\ref{sect:GR} we describe
transformation of the local spectra into the spectra seen by
a distant observer. Examples of interpretation of observed
spectra with the use of theoretical models of partially
ionized, strongly magnetized NS atmospheres are considered
in Sect.~\ref{sect:obs}. In sect.~\ref{sec:concl}, we give
brief conclusions. In the Appendix we briefly describe the effects
of thermal motion of atoms in a strong magnetic field on the
atomic quantum-mechanical characteristics and on the
ionization equilibrium of plasmas, that underlie
calculations of the opacities for the strongly magnetized NS
atmospheres.

\section{The families of Isolated Neutron Stars}
\label{sect:INSfamilies}

A short account of the main properties of the different classes of INSs,
with a focus on their emission in the X-ray range, is useful to set the context
for the observational panorama of thermal emitters.

\begin{figure}[t]
\begin{center}
\includegraphics[height=\linewidth,angle=-90]{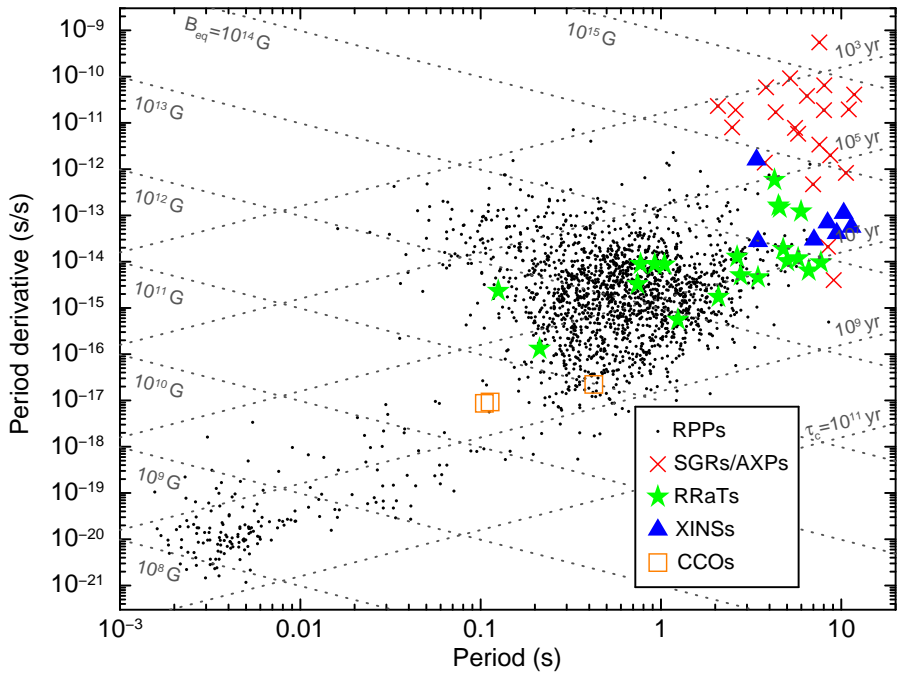}
\caption{$P-\dot{P}$ diagram for the $\sim2300$ currently
known neutron stars with measured $P$ and $\dot{P}$.
Different classes are marked with different symbols.
Lines showing constant $\dot{E}_\mathrm{rot}$
as well as constant inferred surface dipolar magnetic 
field are superimposed.
\label{fig:ppdot}}
\end{center}
\end{figure}

\begin{list}{$\bullet$}{
\setlength\itemindent{3ex}
\setlength\listparindent{1em}
\leftmargin=0ex
}

\item \textbf{Rotation-Powered Pulsars (RPPs)}. 
This is the class of INSs with the largest observational database, more than 2,200
sources being known\footnote{See e.g. the ATNF pulsar
database at http://www.atnf.csiro.au/research/pulsar/psrcat/
\citep{ATNF}}, 
mostly from radio surveys. A large population of radio-silent 
RPPs is also emerging, thanks to gamma-ray observations by the Fermi mission
\citep[see][and references therein]{Abdo13}.

Rotation of a magnetized INS induces electric fields which accelerate particles
in the magnetosphere surrounding the star, initiating electromagnetic cascades. 
This mechanism produces  synchrotron and curvature radiation along a peculiar 
beaming pattern, ultimately related to the magnetic field configuration, at the expense
of the dissipation of the NS rotational energy
\citep[see][for a review]{HardingLai06}. Because of such rotation-powered emission
(see also \citealp{Gwinn_this_volume}, this volume),
RPPs are typically observed as pulsating sources from radio wavelengths\footnote{Indeed, 
their discovery as radio pulsars 
was the first observational evidence 
for the existence of NSs} to very high 
energy gamma-rays -- the spin-down luminosity
$\dot{E}_\mathrm{rot}=4\pi^2 I \dot{P}P^{-3}$ 
(where $I\sim10^{45}$ g cm$^2$ is the NS moment of inertia) being indeed much larger 
than the inferred electromagnetic luminosities. However, as a matter of fact,
most of the spin-down luminosity is carried away by a relativistic particle wind which,
interacting with the surrounding medium, powers bright pulsar wind nebulae, seen as diffuse 
structures in the radio, X-ray, and gamma-ray energy ranges. 

In this picture, assuming the observed $\dot{E}_\mathrm{rot}$
to be due to magnetic dipole radiation yields an estimate of
the surface dipole field at the magnetic equator  of the
star $B_\mathrm{eq}=3.2\times10^{19}(P\dot{P})^{1/2}$ G, as
well as an estimate of the NS age  $\tau_\mathrm{c}=P/(2\dot{P})$, which is called characteristic age. The
first estimate assumes $\sqrt{I_{45}}/(R_6^3\sin\alpha)=1$,
where  $I_{45}=I/10^{45}$ g cm$^2$, $R_6$ is the stellar
radius $R$ in units of $10^6$ cm, and $\alpha$ is the angle
between the magnetic and rotational axes. More
realistic calculations give similar relations between $B$
and $P\dot{P}$ \citep{Spitkovsky06,BeskinIF13}. The second
estimate assumes a birth period much shorter than the
current period $P$, as well as a non-variable magnetic
field. As can be seen in Figure~\ref{fig:ppdot}, the
bulk of RPPs have $P$ in the 0.1\,--\,1 s range and
$B_\mathrm{eq}\sim10^{12}$ G. The separate subclass  of
Millisecond Pulsars (MSPs), accounting for $\sim10\%$ of the
RPP sample,  is supposed to include very old NSs with a
different evolutionary history, having experienced accretion
in a long-lived binary system which eventually spun-up the
NSs to $P\sim$ a few ms and quenched their magnetic field to
$10^8-10^9$ G. For this reason MSPs are also called recycled pulsars
\citep[][and references therein]{Bisno06}. 

More than 120 RPPs have been detected in the soft X-ray energy range
%
%
\citep[see, e.g.,][]{Marelli11,Abdo13},
 and several of them display thermal emission 
from their hot surfaces in addition to 
 the rotation-powered emission.
While a thermal  component related to the NS cooling is
apparent in a few sources, an additional component
with a higher temperature and a smaller emitting area can
also be observed \citep{deluca05}, probably related
to re-heating of the magnetic polar caps by bombardment of magnetospheric particles.
We note that RPPs with the largest inferred magnetic fields ($B_\mathrm{eq} \sim 10^{13}-10^{14}$ G)
are often considered as a separate, ``High-$B$''  class, (HB, see \citealt{ng11} for a review).
HB RPPs show some evidence for a larger thermal luminosity
when compared to RPPs of similar ages 
and can be the link between magnetars and RPPs,
with magnetic field decay playing an important role
for their thermal evolution \citep{AguileraPM08,PonsMG09}.  
At least in one case (PSR~J1846$-$0258, \citealp{Gavriil08}), 
a peculiar time variability, reminiscent of the behaviour of
the anomalous X-ray pulsars (see below), has been seen in a HB RPP.

\item \textbf{Rotating Radio Transients (RRaTs)}. 
Discovered as sources of repeated, bright, short ($\sim\,1$ ms)  radio bursts
occurring at integral multiples of an underlying periodicity, RRaTs were 
initially considered as a new class of INSs \citep{mclaughlin06}. High time resolution radio surveys are unveiling 
new RRaTs (about 70 are currently known -- see \citealp{burke13} for a recent review). 

It has been observed that RRaTs,
as a class, have a larger magnetic field with respect to the bulk of RPPs,
which prompted speculations about possible relationships of such 
sources with magnetars (or with HB RPPs, or with the
Magnificent Seven NSs considered below).
Interestingly enough, 
the most active RRaT (PSR~J1819$-$1458) -- the only source of its class being detected in 
soft X-rays -- has an X-ray phenomenology (as well as a position in the $P-\dot{P}$ plane)
fully similar to the one of the Magnificent Seven,
with a pulsed thermal-like emission and a broad spectral absorption feature 
superimposed to the continuum \citep{Miller11}. 
In any case, it is currently believed that RRaTs are 
not substantially physically different from RPPs -- which also show time variability of the radio pulses --
RRaTs have just a somewhat extreme temporal behaviour. 

\item \textbf{Soft Gamma Repeaters (SGRs)} and \textbf{Anomalous X-ray Pulsars (AXPs)}. 
About two dozen such sources are known \citep{OlausenKaspi14}. 
Originally thought to form two different classes of objects,
they were later shown to form a single SGR/AXP class
\citep{GavriilKW02}.
Detailed reviews 
of their phenomenology and physics are included in other
chapters of this
volume;
see also \citet{Mereghetti08,Mereghetti13,Turolla13}.

The period of SGR/AXPs (in the 2\,--\,12 s range) is much longer
than the one of the bulk of RPPs; their period derivative (showing peculiar time variability)
is also very large, corresponding to dipole magnetic fields
of the order of $B\sim10^{14}-10^{15}$ G
(see Figure~\ref{fig:ppdot}).

These objects show a very complex high energy phenomenology. This 
includes a persistent, pulsed soft X-ray component (often with a transient behavior), 
dominated by thermal-like emission, typically with a luminosity largely exceeding the spin-down 
luminosity. Moreover, SGR/AXPs display
a spectacular flaring emission component \citep[see][for a
review]{Rea11}, best seen from a few keV to a few hundreds keV. 
They are typically radio silent, although very bright radio pulses have been 
observed from a few sources.  

SGR/AXPs are generally believed to be magnetars 
(see the review by \linebreak\citealp{OlausenKaspi14}),
i.e.{} NSs ultimately
powered by the decay of their magnetic field
\citep{DuncanThompson92}.\footnote{There are alternative 
hypotheses about the nature of SGR/AXPs, assuming their braking
by an accretion disk 
\citep[e.g.,][and references therein]{Truemper-ea13}
or a ``magnetic slab'' \citep{BisnoIkhsanov14}, 
as well as the models of drift waves in the magnetosphere of
a neutron star with $B\sim10^{12}$~G \citep{Malov10}
or rapidly rotating massive ($M>M_\odot$) white dwarfs with
$B\sim10^8$~G \citep{Boshkaev-ea13}.} 
The magnetic field of magnetars is supposed to be much more
intense than for other classes of INSs and to have a peculiar topology.  

\item \textbf{Central Compact Objects (CCOs)}.
This class includes a dozen of point sources, lying close to
the center of young (0.3\,--\,10 kyr) Supernova Remnants, 
discovered in the soft X-ray band and supposed to be young INSs \citep{Deluca08,GotthelfHA13}. 
They have no counterparts at any other wavelength,
nor associated diffuse nebulae, nor any manifestation of rotation-powered magnetospheric activity.
Such sources only display steady, thermal-like X-ray 
emission, typically with a two-tem\-pe\-r\-a\-ture blackbody spectrum and very small emitting areas.
Absorption features superimposed to the continuum have been observed in two sources.

Measurement of a tiny period derivative for the only three pulsating CCOs
point to very low surface dipole fields ($B_\mathrm{eq}\sim10^{11}$ G, \citealp{GotthelfHA13}).
This is consistent with an interpretation of the absorption lines as electron cyclotron features. 
In this picture, it is not clear what is the origin of the apparent, 
relatively large thermal luminosities and large temperature anisotropies for these NSs. 
The spin-down luminosity is too low to explain the hot spot emission, while
X-ray timing, as well as deep optical/infrared observations, rule out
the existence of companion stars as well as accretion from debris discs 
\citep{HalpernGotthelf,Deluca11}. It has been suggested that 
a strong crustal toroidal magnetic field could channel heat from the star interior, producing 
hot spots on the surfaces, while not influencing the NS
spin-down \citep[see, e.g.,][]{Shabaltas12}. 

As another puzzle, the inferred, relatively high birth rate of CCOs clashes with the apparently
underpopulated region of the $P-\dot{P}$ diagram in which they reside
(\citealp{Kaspi10,HalpernGotthelf}
-- see Figure~\ref{fig:ppdot}).
It has been proposed that the weak observed dipole field could result from prompt accretion
of $10^{-4}-10^{-3}\,M_{\odot}$ fallback material soon after the SN explosion. The buried field would re-emerge 
later, on a $10^3-10^5$ yr time scale (depending on the amount of accreted material), turning a CCO into a RPP, 
or even into a magnetar \citep{Vigano12}.

\item \textbf{The Magnificent Seven, or X-ray INSs
(XINSs)}.
These seven INSs were discovered in the 
\textit{ROSAT All Sky Survey}. Reviews on their properties  
are given by \citet{haberl07,turolla09,kaplan09b},
to whom the reader is referred for more details and references.

The XINSs have spinning periods in the same range as SGR/AXPs, but the period derivative
points to a dipole field $\sim1$ order of magnitude lower
for the XINSs (but still well above 
the one of the bulk of RPPs
-- see Figure~\ref{fig:ppdot}).

These sources display a thermal-like spectrum,  from
the optical to the soft X-ray range.\footnote{For some of
them, pulsed radio emission has been also detected at very
low frequency 111 MHz at the the Pushchino Radio
Astronomy Observatory -- see 
\citet{MalofeevMT07,Teplykh-ea11}. Such detections are very
intriguing and await confirmation from other observations at
similar frequencies, e.g., by the LOFAR telescope.}
Their thermal luminosity is somewhat larger than expected
from conventional cooling of NSs at the same characteristic
age. Broad (often multiple) spectral distortions compatible with
absorption features are seen, superimposed on the thermal continuum. 
The nature of these features is unclear. Suggested explanations include, 
for example, atomic  or proton-cyclotron features  \citep[e.g.,][]{haberl07},
photoionization in a relatively dense cloud in the vicinity
of the NS \citep{Hambaryan-ea09}, or a result of a complex
strongly inhomogeneous distribution of temperature over the
surface \citep{vigano14}. In one case, the observed
spectral shape has been tentatively explained as originated
in a thin partially-ionized hydrogen atmosphere above a
condensed iron surface (see Sect.~\ref{sect:RBS1223}).

All XINSs have an optical/UV counterpart, with a flux exceeding by a factor 5\,--\,50 (around
5,000 \AA) the expected value, based on the extrapolation of the soft X-ray spectrum \citep{kaplan11}. 
Such optical/UV excesses show power-law spectra, but the observed slopes
differ from source to source and are generally not consistent with 
the Rayleigh-Jeans tail of a blackbody spectrum. The origin of the
optical/UV excess is not understood (atmospheric effects? magnetospheric activity?
evidence for an emitting region larger and cooler than the one seen in X-rays?).

All XINSs are steady emitters, with the exception of RX~J0720.4$-$3125, whose 
time variability has not yet been understood (see \citealp{hohle12}, 
possible explanations range from magnetar-like activity to a peculiar accretion episode).

The XINSs are rather close (from 120 pc to a few hundred pc)
to the Solar system. These distance estimates are based on direct annual parallax measurements
or on comparison of the observed photoelectric absorption to models
for the 3D distribution of local interstellar medium, and
they are consistent with the 
large observed proper motions. Back-projection of the NS space trajectories allowed to infer 
a kinematic age for a few sources, based on possible association with clusters of massive stars
\citep[see e.g.][]{Tezlaff11}.
Interestingly enough, the kinematic ages are systematically
smaller than characteristic ages $\tau_\mathrm{c}$, being
thus more consistent with the relatively high thermal
luminosities.

It has been suggested that the magnetic field of the XINSs
could have decayed since their birth,
affecting their rotational and thermal evolution
\citep{popov10}. The relationship of the XINSs to other 
classes of INSs is in any case still poorly understood (aged
magnetars? HB RPPs  with unfavorable radio beaming? extreme
RRaTs?).

\end{list}

Unifying the apparent diversity 
of the classes of INSs in a coherent physical scenario 
is a major goal in the astrophysics of INSs. 
A study in this perspective was performed
by \citet{vigano13}, who modeled 
the coupled evolution of temperature and 
magnetic field (driving the rotational 
evolution) of INSs and proved the possibility
to explain the overall properties of the classes
of SGR/AXPs, HB RPPs, and of the XINSs by varying the 
initial magnetic field, mass and envelope composition
in a unique parent population of INSs.

\section{Isolated neutron stars with thermal spectra}
\label{sec:objects}

\begin{table*}
\footnotesize
\setlength{\tabcolsep}{4pt}
\begin{center}
\begin{tabular}{l c c c c c c c c}
\hline
\hline
Source  & $T_\mathrm{bb}$ & $R_\mathrm{bb}$ &
$T_\mathrm{nsa/rcs}$ & $R_\mathrm{nsa}$ &  best fit &
$T_\mathrm{cool}$ & $\log(L_\mathrm{cool})$ \\
  & [$10^6$ K] & [km] & [$10^6$ K] & [km]  & model & [$10^6$ K] & [erg s$^{-1}$] \\
\hline

CXOU J185238.6+004020   & 5.1 &  0.9 & 3.37 & 3.0 &  BB/nsa     &  $< 1.1$  &  $<33.1$ \\
1E 1207.4$-$5209**      & 2.2 &  9.6 & 1.7  & 7.4 &  BB*/nsa*   &  $<  0.7$ &  $<32.2$ \\
RX J0822$-$4300         & 4.6 &  1.7 & 2.37 & 6.4 &  BB/nsa     &  $<  1.0$ &  $<32.9$ \\
CXO J232327.9+584842    & 5.2 &  1.7 & 3.34 & 2.7 &  BB/nsa     &  $< 1.3$  &  $<33.3$ \\
1WGA J1713.4$-$3949     & 4.8 &  0.4 & --   &  -- &  BB+PL      &     --    &     --   \\
CXOU J085201.4$-$461753 & 4.7 & 0.28 & 3.1  & 1.2 &  BB/nsa     &     --    &     --   \\
XMMU J172054.5$-$372652 & 4.9 &  2.6 &  --  &  -- &  BB+PL      &     --    &     --   \\
XMMU J173203.3$-$344518 & 5.7 &  --  & 2.6$^\mathrm{C}$  & 13$^C$  &  C atm & -- & --  \\
PSR J0538+2817          & 1.9 &  2.6 &  --  &  -- &  BB+PL      & $<  0.6$  &  $<31.9$ \\
PSR B1055$-$52          & 2.2 &  0.3 &  --  &  -- & 2BB+PL      &     0.8   &      --  \\
PSR J0633+1746          & 1.6 &  0.1 &  --  &  -- & 2BB+PL      &    0.49   &      --  \\
PSR B1706$-$44          & 1.7 &  3.3 &  --  &  -- &  BB+PL      &  $< 0.7$  &  $<32.2$ \\
PSR B0833$-$45          & 1.4 &  5.0 & 0.93 & 9.4 & (BB/nsa)+PL &  $< 0.5$  &  $<31.5$ \\
PSR B0656+14            & 1.2 &  2.4 &  --  &  -- & 2BB+PL      &    0.6    &      --  \\
PSR B2334+61            & 1.9 &  1.1 & 1.0  & 7.9 &  BB/nsa     &  $< 0.6$  &  $<31.9$ \\
PSR J1740+1000          & 2.0 &  0.4 & 0.6  &10.3 & 2BB/nsa     &    0.9    &      --  \\
PSR J1741$-$2054        & 0.7 &  12  &  --  &  -- &  BB         &   --      &      --  \\
PSR J1357$-$6429        & 1.6 &  2.0 & 0.74 &10.0 & (BB/nsa)+PL &   --      &      --  \\
PSR J0726$-$2612        & 1.0 &  4.6 &  --  &  -- &  BB         & $< 0.46$  &  $<31.5$ \\
PSR J1119$-$6127**      & 3.1 &  1.5 &  --  &  -- &  BB         &  $< 1.4$  &  $<32.9$ \\
PSR J1819$-$1458        & 1.5 & 12.3 &  --  &  -- &  BB         &     --    &      --  \\
PSR J1718$-$3718        & 2.2 &  2.0 &  --  &  -- &  BB         &  $<  1.0$ &  $<32.9$ \\
RX J0420.0$-$5022       & 0.6 &  3.4 &  --  &  -- &  BB         &     --    &      --  \\
RX J1856.5$-$3754**     & 0.73&  4.1 &  --  &  -- &  BB         &     --    &      --  \\
RX J2143.0+0654         & 1.24&  2.3 &  --  &  -- &  BB         &     --    &      --  \\
RX J0720.4$-$3125       & 0.97&  5.7 &  --  &  -- &  BB         &     --    &      --  \\
RX J0806.4$-$4123       & 1.17&  1.2 & 0.63 & 8.2 &  BB*/nsa*   &     --    &      --  \\
RX J1308.6+2127**       & 1.09&  5.0 &  --  &  -- &  BB*        &     --    &      --  \\
RX J1605.3+3249         & 1.15&  0.9 & 0.49 & 7.0 &  BB*/nsa*   &     --    &      --  \\

\hline
\hline
\end{tabular}
\end{center}
{}* Absorption line(s) \texttt{gabs} included in the fit.\\
{}** See discussion of a more elaborated analysis in Sect.~\ref{sect:obs}.\\
$^\mathrm{C}$ Fits to a carbon atmosphere and assuming $d=3.2$ kpc. \\
\caption{Emission properties of the thermally emitting neutron
stars.  $T_\mathrm{bb}$ and $R_\mathrm{bb}$
are the temperature and radius inferred by the \texttt{bbodyrad} model.
$T_\mathrm{nsa}$ is the temperature inferred by the
\texttt{nsa} model 
with acceptable
associated radius $R_\mathrm{nsa}$, also indicated.
$T_\mathrm{cool}$ is either the lower temperature
for models including 2 BB, compatible with emission from the entire
surface, or the upper limit for cases showing emission from a small spot
$R_\mathrm{bb}\sim$ a few km. In the latter case, $L_\mathrm{cool}$ is 
the associated upper
limit to the ``hidden`` thermal luminosity.  Data are from 
\citet{vigano13} as well as from
www.neutronstarcooling.info. All radii, temperatures and luminosities 
are the values as measured by a distant observer.
}
\label{tab:spectral}
\end{table*}

\citet{vigano13} gathered and thoroughly re-analyzed all the best available data
on isolated, thermally emitting NSs were in a consistent way.
We refer the interested reader to that reference for more details.
The data sample of 40 sources was compared to theoretical models of the magneto-thermal evolution 
of NSs, in an attempt to explain the phenomenological diversity of SGR/AXPs, high-B radio-pulsars, and isolated nearby
NSs by only varying their initial magnetic field, mass and envelope composition. The cooling theory of
NSs and several issues characteristic of magnetars are also
discussed in other chapters of this
volume. In this chapter, we focus on the thermal emission from
sources not cataloged as magnetar candidates (SGR/AXPs).
The sample of selected sources (see Table\,\ref{tab:spectral}) includes:
\begin{list}{$\bullet$}{
\setlength\itemindent{3ex}
\setlength\listparindent{1em}
\leftmargin=0ex
}
\item
 Eight CCOs, including the very young NS in Cassiopeia A,
and the only three CCOs with measured values of $P$ and $\dot{P}$. 
We have ignored the other CCOs
candidates, since they have spectral information with poor
statistics and/or a very uncertain age of the associated SNR.

\item
13 rotation powered pulsars, including the Vela pulsar and
the so-called Three Musketeers (PSR~B0656, PSR~B1055 and the
$\gamma$-ray-loud Geminga; \citealt{deluca05}). We have
excluded most of the young pulsars, many of which are
associated with pulsar wind nebulae, 
since in those cases data are compatible with
non-thermal emission powered by the rotational energy loss,
which is orders of magnitude larger than their X-ray
luminosity (i.e. RX J0007.0+7303 in SNR CTA1;
\citealt{caraveo10}). We also exclude several old  pulsars
\citep{zavlin04} with thermal emission from a tiny hot spot
(a few tens of m$^2$), since the  temperature of the small
hot spots is probably unrelated to the cooling history of
the NS. Our list  also includes four high-B radio-pulsars 
magnetic fields $B_\mathrm{eq} \sim 10^{13}-10^{14}$ G and
good quality spectra.  We have excluded the AXP-like pulsar
PSR~J1846$-$0258 since during quiescence its X-ray emission
does not show a significant thermal component
\citep{ng08,livingstone11b}, and it is orders of magnitude
smaller than its rotational energy loss.

\item
We have included the only RRaT detected so far
in X-ray (PSR~J1819$-$1458).

\item
 The Magnificent Seven (XINSs).
All of them have good spectra,
and most of them have well determined timing properties and good
distance determinations (sometimes with direct parallax
measurements).
\end{list}

We summarize the main properties of thermal emitters in Table \ref{tab:spectral} (taken from 
\citealp{vigano13} and extended).
All the data presented here with links to abundant references can also be found in the
website {\rm http://www.neutronstarcooling.info/}. 


Although both, luminosities and temperatures can be obtained by spectral analysis, 
it is usually difficult to determine them accurately. The
luminosity is always subject to the uncertainty in the distance
measurement, while the inferred effective temperature depends on the choice
of the emission model (blackbody vs. atmosphere models, composition,
condensed surface, etc.), and it carries large theoretical uncertainties in
the case of strong magnetic fields.  We often find that more than one
model can fit equally well the data, without any clear, physically
motivated preference for one of them.  Photoelectric
absorption from interstellar medium further constitutes a source of
error in temperature measurements, since the value of the hydrogen
column density $N_\mathrm{H}$ is correlated to the temperature value obtained
in spectral fits. Different choices for the absorption model and the
metal abundances can also yield different results for the temperature.
In addition, in the very common case of the presence of inhomogeneous
surface temperature distributions, only an approximation with two or
three regions at different temperatures is usually employed.
Moreover, in the case of data with few photons and/or strong absorption features,
the temperature is poorly constrained by the fit, adding a large
statistical error to the systematic one. 

\section{Atoms and matter in strong magnetic fields}
\label{sect:field-effects}

A general review of the physics of matter in strong magnetic
fields is given by \citet{Lai_this_volume} in this volume. 
Here we consider only the concepts that are crucial for the
theory of formation of thermal spectra of NSs with
strong magnetic fields.

\subsection{Landau quantization}
\label{sect:QLandau}

It is convenient to express the magnetic field by its strength in atomic units, $\gamma$, or
in relativistic units, $b$:
\begin{equation} 
\gamma = B/B_0 = 425.44\,B_{12},
\qquad
 b =
{\hbar\omc}/({\mel c^2}) = B/B_\mathrm{QED} =
{B_{12}}/{44.14} \,.
\label{magpar}
\end{equation}
Here, 
$
   B_0 = {\mel^2\,c\,e^3}/{\hbar^3}
$
is the atomic unit of magnetic field,
$B_{12}\equiv B/10^{12}$~G, $\omc=eB/\mel c$ is the
electron cyclotron frequency, and $B_\mathrm{QED}
= \mel^2 c^3 / (e\hbar) = B_0/\alphaf^2$ is the critical
(Schwinger) field, above which specific effects of quantum
electrodynamics (QED)
become pronounced. We call a
magnetic field \emph{strong}, if $\gamma\gg1$, and
\emph{superstrong}, if $b\gtrsim1$.

The motion of charged particles in a magnetic field $\bm{B}$ is quantized
in discrete Landau levels, whereas the longitudinal
(parallel to $\bm{B}$) momentum of the particle can change
continuously.
In the nonrelativistic theory, the threshold excitation
energy of the $N$th Landau level is $N\hbar\omc$ ($N=0,1,2,\ldots$). In the
relativistic theory, it is $E_N=\mel c^2
\,(\sqrt{1+2bN}-1)$. The wave functions
that describe an electron in a magnetic field \citep{SokTer}
have a characteristic transverse scale of the order of the
``magnetic length'' $\am=(\hbar
c/eB)^{1/2}=\aB/\sqrt{\gamma}$, where $\aB$ is the Bohr
radius. The momentum projection on the magnetic field
remains a good quantum number, therefore 
these projections have the usual Maxwellian distribution
at thermodynamic equilibrium. For transverse motion,
however, we have the discrete Boltzmann distribution over
$N$.

In practice, Landau quantization becomes important when
the electron cyclotron energy $\hbar\omc$ is at least
comparable to both the electron Fermi energy $\EF$ and
temperature $T$ (in  energy units, i.e., $10^6\mbox{
K}=86.17$ eV). If $\hbar\omc$ is
appreciably larger than both these energies, then most
electrons reside on the ground Landau level in thermodynamic
equilibrium, and the field is called \emph{strongly
quantizing}.
It is the case, if conditions 
$\rho<\rho_B$ and $\zete\gg1$ are fulfilled  simultaneously, where
\begin{equation}
  \rho_B =
 \frac{\mion}{\pi^2\sqrt2\,\am^3\,Z}
  = 7045 \,\frac{A}{Z}
       \,B_{12}^{3/2}\text{ \gcc},
\qquad
   \zete = \frac{\hbar\omc}{T} = 134.34\,
   \frac{B_{12}}{T_6} ,
\label{zeta_e}
\end{equation}
 $\mion=Am_\mathrm{u}$ is the ion
mass, $m_\mathrm{u}$ is the unified atomic mass unit,
and $T_6$ is temperature in units of $10^6$~K.
In NS atmospheres, these conditions are
satisfied, as a rule, at  $B\gtrsim10^{11}$~G. In the
opposite limit $\zete\ll1$, the field can be considered as
\emph{nonquantizing}. In the magnetospheres, which have lower
densities, electrons can condensate on the lowest Landau
level even at $B\sim10^8$~G because of the violation of the
LTE conditions \citep[e.g.,][]{Mesz}, but this is not the
case in the photospheres (the atmospheric 
layers whose thermal state is determined by the radiative
flux and where the observed spectrum is mainly formed;
\citep[see][]{P14}).

For ions, the
parameter $\zete$ is replaced by 
\begin{equation}
   \zeti = \hbar\omci/T = 0.0737\,(Z/A)
           B_{12}/T_6.
\label{omci}
\end{equation}
Here, $\omci=ZeB/(\mion c)$ is the ion cyclotron frequency,
$Ze$ is the ion charge,
and $\hbar\omci=6.35(Z/A)B_{12}$~eV is the ion cyclotron
energy. In magnetar atmospheres, where $B_{12}\gtrsim100$
and $T_6\lesssim10$, the parameter $\zeti$ is not small,
therefore the quantization of the ion motion should be
taken into account. A parameter analogous to $\rho_B$ is
unimportant for ions, because they are nondegenerate in
NS envelopes.

\subsection{Bound species in strong magnetic fields}
\label{sect:atoms}

As first noticed by \citet{CLR70}, atoms with bound states
should be much more abundant at $\gamma\gg1$ than at
$\gamma\lesssim1$ in a NS atmosphere at the same
temperature. This difference is caused by the
magnetically-induced increase of binding energies (and
decrease of size) of atoms in so-called \emph{tightly-bound
states}, which are characterized by electron-charge
concentration at short distances to the nucleus. Therefore
it is important to consider the bound states and bound-bound
and bound-free transitions in a strong magnetic field even
for light-element atmospheres, which would be almost fully
ionized at $T\sim10^5$~K in the nonmagnetic case.

Most studies of atoms in strong magnetic fields
have considered an atom with an infinitely heavy (fixed in
space) nucleus (see, e.g., \citealt{Garstang77,Ruder-ea},
for reviews). This model is rather crude, but it is a
convenient first approximation. In this section
we review this model. An outline of more accurate
treatments, which take the effects of finite atomic mass
into account, is given in Appendix~\ref{sect:motion}.

\subsubsection{One-electron atoms and ions}
\label{sect:Hlike}

The H atom in a magnetic field is well studied. At $B>10^9$~G, its only electron 
resides at the ground Landau level $N=0$. Since $N$ is fixed, the quantum
state is determined by two other quantum numbers:
$s = 0,1,2,\ldots$, which corresponds to the electron
angular-momentum projection on the magnetic-field direction,
$-\hbar s$, and $\nu=0,1,2,\ldots$, which corresponds mainly
to the motion along $\bm{B}$.
The tightly-bound states all have $\nu=0$. Their binding energies
logarithmically increase with increasing $\gamma$
(asymptotically as $\sim\ln^2\gamma$~Ry, where 1~Ry=13.6057 eV is the
Rydberg constant in energy units). Non-zero
values of $\nu$ correspond to loosely-bound states, whose
binding energies are confined within 1~Ry.

Accurate calculations of the properties of the bound states
of a non-moving H atom in a strong magnetic field have been
performed in many works (see \citealt{Ruder-ea}, for a
review).  The $B$-dependences of binding energies are well
approximated by analytical functions \citep{P14}.  Continuum
wave functions and photoionization cross-sections have also
been calculated \citep{PPV97}.

In the approximation of an infinite nuclear mass, the energy
of any one-electron ion is related to the H-atom
energy as $E(\Znuc,B)=\Znuc^2\,E(1,B/\Znuc^2)$
\linebreak\citep{SurmelianOConnel74}. Analogous
similarity relations exist also for the cross sections of
radiative transitions \citep{Wunner-ea82}. However, all
these relations are
violated by motion across the magnetic field. Even for an
atom at rest, the account of the finite nuclear mass can be
important at $s\neq0$ (see Appendix ~\ref{sect:motion}).

\subsubsection{Many-electron atoms and ions}

According to the Thomas-Fermi model, a typical size of an
atom with a large nuclear charge $\Znuc$ is proportional
to $\gamma^{-2/5}$ in the interval $\Znuc^{4/3} \ll \gamma
\ll \Znuc^3$ \citep{Kadomtsev70}, but this model breaks down at $\gamma\gtrsim\Znuc^3$
\citep{LiebSY92}. In particular, it cannot describe the
difference of the transverse and longitudinal atomic sizes,
which becomes huge in such strong fields. In this
case, however, a good starting approximation is the so
called adiabatic approximation, which presents each electron
orbital as a product of the Landau function that describes
free electron motion transverse to the field
\citep{SokTer} and a function describing a one-dimensional
motion along $\bm{B}$ in an effective potential,
similar to a truncated one-dimensional Coulomb potential
\citep{HainesRoberts}. At $\gamma\gg\Znuc^3$, all electron shells of the atom are
strongly compressed in the directions transverse to the
field. In the ground state, atomic sizes along and
transverse to $\bm{B}$ can be estimated as
\citep{KadomtsevKudryavtsev}
\begin{equation}
l_\perp\approx \sqrt{2\Znuc-1}\,\am,
\quad
l_z\approx\frac{\Znuc^{-1}\aB
}{
\ln[\sqrt{\gamma}/(\Znuc\sqrt{2\Znuc-1})]}.
\end{equation}
In this case, the binding energy of the ground
state increases with increasing $\bm{B}$ asymptotically as 
$E^{(0)}\sim -\Znuc\hbar^2/(\mel l_z^2)$
\citep{KadomtsevKudryavtsev}.

Thomas-Fermi results are useful as an
order-of-magnitude estimate. More accurate calculations
of binding energies and oscillator strengths of
many-electron atoms were performed with different methods. 
For atoms with small nuclear charge numbers $\Znuc$, such as
helium, a sufficiently accurate and practical method is the
Hartree-Fock method with trial orbitals in the adiabatic
approximation \citep{Ruder-ea,MillerNeuhauser91,MedinLP08}.
But the
condition of applicability of the adiabatic approximation
$\gamma\gg\Znuc^3$ is too restrictive for larger $\Znuc$. 
It is overcome in the
mesh Hartree-Fock method, where each orbital is
numerically determined as a function of the longitudinal
($z$) and radial coordinates 
(e.g., \citealp{IvanovSchmelcher00}, and
references therein).
This method, however, is computationally expensive.
\citet{MoriHailey02} proposed a ``hybrid''
method, where corrections to the adiabatic Hartree
approximation due to electron exchange and admixture of
higher Landau levels are treated as perturbations. The latter method
proved to be practical for modeling NS atmospheres
containing atoms and ions of elements with  $2<
\Znuc\lesssim10$, because it can provide an acceptable
accuracy  at moderate computational expenses.

\subsubsection{Molecules and molecular ions}

Best studied molecules and molecular ions are diatomic systems,
especially the H$_2$$^+$ ion (\citealp{KS96}, and references
therein) and the H$_2$ molecule \linebreak(\citealp{SchmelcherDC01}, and references
therein). \citet{Lai01} obtained approximate expressions for
binding energies of low-lying levels of the H$_2$  molecule
at $\gamma\gtrsim10^3$. These energies increase
approximately at the same rate $\propto(\ln\gamma)^2$ as the
binding energies of tightly-bound states of the atom. In
such strong fields, the ground state of the H$_2$ molecule
is the state where the spins of both electrons are
counter-aligned to $\bm{B}$ and the molecular axis is parallel to
$\bm{B}$, unlike the weak fields where the ground state is
$^1\mathrm{\Sigma}_g$.

In moderately strong fields (with $\gamma\sim1-10$), the
behavior of the molecular terms is complicated. If the H$_2$
molecule is oriented along $\bm{B}$, then its states
$^1\mathrm{\Sigma}_g$ and $^3\mathrm{\Pi}_u$ are metastable
at $0.18<\gamma<12.3$ and decay into  $^3\mathrm{\Sigma}_u$,
which is unbound \citep{DetmerSC98}. It turns out, however,
that the orientation along $\bm{B}$ is not optimal in this
case. For example, the lowest energy is provided by the
orientation of the molecule in the triplet state at
90$^\circ$ to $\bm{B}$ at $\gamma=1$, and 37$^\circ$ at
$\gamma=10$ \citep{Kubo07}.

Strong magnetic fields stabilize the He$_2$  molecule and its
ions He$_2$$^{+}$, He$_2$$^{2+}$, and He$_2$$^{3+}$, which
do not exist in the absence of the field. \citet{MoriHeyl}
have performed the most complete study of their binding
energies in NS atmospheres. The ions HeH$^{++}$,
H$_3$$^{++}$, and other exotic molecular ions, which become
stable in strong magnetic fields, were also considered
(see \citealt{Turbiner07,TurbinerLVG10}, and references
therein). An evaluation of the ionization equilibrium
shows that, at densities, temperatures, and magnetic fields characteristic of NSs,
the abundance of such ions (as well as that of H$_2$$^+$ ions considered by
\citealp{Khers87b}) is too small to affect the thermal spectrum.

There are very few results on molecules composed of atoms
heavier than He. In particular, \citet{MedinLai06a}
applied the density-functional method to
calculations of binding energies of various molecules from
H$_n$ to Fe$_n$ with $n$ from 1 through 8 at $B$ from
$10^{12}$~G to $2\times10^{15}$~G. The earlier studies of
heavy molecules in strong magnetic fields are discussed in
the review by \citet{Lai01}.

\subsection{Atmosphere thermodynamics}

According to the Bohr-van Leeuwen theorem, the
magnetic field does not affect thermodynamics of classical
charged particles. The situation differs in quantum
mechanics. The importance of the quantum effects depends on
the parameters $\zete$ and $\zeti$
[Eqs.~(\ref{zeta_e}), (\ref{omci})].

Studies of thermodynamics of magnetic NS
atmospheres, as a rule, are based on the
decomposition of the Helmholtz free energy
\begin{equation}
   F= F_\mathrm{id}^\mathrm{(e)} + F_\mathrm{id}^\mathrm{(i)} +
      F_\mathrm{int} +
      F_\mathrm{ex},
\label{F}
\end{equation}
where $F_\mathrm{id}^\mathrm{(e)}$ and $F_\mathrm{id}^\mathrm{(i)}$
describe the ideal electron and ion gases, $F_\mathrm{int}$
includes internal degrees of freedom for bound states, and
$F_\mathrm{ex}$ is a nonideal component due to interactions
between plasma particles. All the necessary thermodynamic
functions are then expressed through
derivatives of $F$ over $\rho$ and $T$.

\subsubsection{Equation of state}
\label{sect:EOS}

The free energy of $N_\mathrm{i}$
nondegenerate nonrelativistic ions is given by
\begin{equation}
 \frac{F_\mathrm{id}^\mathrm{(i)}}{N_\mathrm{i}T} =
      \ln\left(2\pi \frac{\nion \lambdi\am^2}{Z}\right)
    + \ln\left( 1 - \mathrm{e}^{- \zeti}\right) - 1
      + \frac{\zeti}{2} +  \ln\Bigg(\frac{
     \sinh[\gfact\,\zeti \Mspin/4] }{ \sinh(\gfact\,\zeti/ 4)
       }
     \Bigg),
\label{Fp}
\end{equation}
where $\lambdi = [2\pi\hbar^2 / (\mion T)]^{1/2}$ is the
thermal de Broglie wavelength for the ions, $\sion$ is the
spin number, and $\gfact$ is the spin-related g-factor (for
instance, $\sion=1/2$ and $\gfact=5.5857$ for the proton). All the
terms in  (\ref{Fp}) have clear physical meanings. At
$\zeti\to0$, the first and second terms give together $\ln(
\nion \lambdi^3)$, which corresponds to the
three-dimensional Boltzmann gas. The first term corresponds
to the one-dimensional Boltzmann gas model at $\zeti\gg1$.
The second-last term in (\ref{Fp}) gives the energy
$\hbar\omci/2$ of zero-point oscillations of every ion transverse to
the magnetic field. Finally, the last term represents the
energy of magnetic moments in a magnetic field.

The ideal electron-gas part of the free energy
$F_\mathrm{id}^\mathrm{(e)}$ can be
expressed through the Fermi-Dirac integrals \citep[see][for
explicit expressions]{PC13}. 
In a strongly quantizing magnetic field, 
the electron Fermi momentum  equals $\pF =
2\pi^2\am^2\hbar n_\mathrm{e}$, where $n_\mathrm{e}=\nion Z$
is the electron number density. Therefore, with increasing
$n_\mathrm{e}$ at a fixed $B$, the degenerate electrons
begin to fill the first Landau level when $n_\mathrm{e}$
reaches $n_B=(\pi^2\sqrt2\,\am^3)^{-1}$. This value just
corresponds to the density $\rho_B$ in \req{zeta_e}. The
ratio of the Fermi momentum $\pF$ in the strongly quantizing
field to its nonmagnetic value $\hbar(3\pi^2
n_\mathrm{e})^{1/3}$ equals
$[{4\rho^2}/{(3\rho_B^2)}]^{1/3}$. Therefore, the Fermi
energy at a given density $\rho<\sqrt{3/4}\,\rho_B$ becomes
smaller with increasing $B$, which means that a strongly quantizing
magnetic field relieves the electron-gas degeneracy. For
this reason, strongly magnetized NS atmospheres
remain mostly nondegenerate, despite their densities are orders of
magnitude higher than the nonmagnetic atmosphere densities.
For the nondegenerate electron gas,
$F_\mathrm{id}^\mathrm{(e)}$ takes the form of \req{Fp}
(with the obvious replacements of $\nion$, $\lambdi$,
$\zeti$, $\gfact$, $\sion$, and $Z$ by $n_\mathrm{e}$,
$\lambda_\mathrm{e}$, $\zete$, 2, $\frac12$, and 1,
respectively).

The nonideal free-energy part $F_\mathrm{ex}$ contains the
Coulomb and exchange contributions of the electrons and the
ions, and the electron-ion polarization energy, and
also interactions of ions and electrons with
atoms and molecules. In turn, the interaction between the
ions is described differently depending on the phase state
of matter. The terms that constitute $F_\mathrm{ex}$ depend
on magnetic field only if it quantizes the motion of these
interacting particles. Here we will not discuss these terms
but address an interested reader to \citet{PC13}
and references therein. This nonideality is usually
negligible in the NS atmospheres, but it
determines the formation of a condensed surface, which will
be considered in Sect.\,\ref{sect:cond}.

\subsection{Ionization equilibrium}

\label{sect:Saha}

For atmosphere simulations, it is necessary to determine
the fractions of different bound states, because they affect
the spectral features that are caused by bound-bound and
bound-free transitions. The solution to this problem is
laborious and ambiguous. The principal difficulty in the
chemical model of plasmas, namely the necessity to distinguish
the bound and free electrons and ``attribute'' the bound
electrons to certain nuclei, becomes especially acute at
high densities, where the atomic sizes cannot be anymore
neglected with respect to their distances. Current
approaches to the solution of this problem are based, as a
rule, on the concept of the so-called occupation probabilities
of quantum states. 

Let us consider a quantum state $\kappa$
of an ion lacking $j$ electrons, with binding
energy  $E_{j,\kappa}$ and quantum statistical weight
$g_{j,\kappa}$. An occupation probability $w_{j,\kappa}$
is an additional statistical weight of this quantum state,
caused by interactions with surrounding plasma (in general,
this weight is not necessarily less than unity, therefore it
is not quite a probability). As first noted by
\citet{Fermi24}, occupation probabilities
$w_{j,\kappa}$ cannot be arbitrary but should be consistent
with $F_\mathrm{ex}$. Minimizing $F$ with account of the
Landau quantization leads to a system of
ionization-equilibrium equations for $n_j\equiv \sum_\kappa
n_{j,\kappa}$ \citep[e.g.,][]{RRM97}
\begin{equation}
 \frac{n_j}{n_{j+1}}
      = {n_\mathrm{e} \lambda_\mathrm{e}^3}\,
      \frac{\sinh(\zeta_j/2)}{\zeta_j}
         \,\frac{\zeta_{j+1}}{\sinh(\zeta_{j+1}/2)}
          \,\frac{\tanh(\zete/2)}{\zete} \,
          \frac{\mathcal{Z}_{\mathrm{int},j}}{
            \mathcal{Z}_{\mathrm{int},j+1}}\,
           \exp\left(\frac{E_{j,\mathrm{ion}}}{T}\right),
\label{Saha-m}
\end{equation}
where $\mathcal{Z}_{\mathrm{int},j}=\sum_\kappa
g_{j,\kappa}\,w_{j,\kappa}\,
\exp\left[({E_{j,\kappa}-E_{j,\mathrm{gr.st}} })/({ 
T})\right]$ is internal partition function for the $j$th
ion type,  $E_{j,\mathrm{gr.st}}$ is its ground-state
binding energy,
$E_{j,\mathrm{ion}}=E_{j,\mathrm{gr.st}}-E_{j+1,\mathrm{gr.st}}$
is its ionization energy, and $\zeta_j$ is the magnetic
quantization parameter (\ref{omci}). Equation (\ref{Saha-m})
differs from the usual Saha equation, first, by the terms
with $\zete$ and $\zeta_j$, representing partition
functions for distributions of free electrons and ions over
the Landau levels, and second, by the occupation
probabilities $w_{j,\kappa}$ in the expressions for the
partition functions $\mathcal{Z}_{\mathrm{int},j}$.
Here, $w_{j,\kappa}$ are the thermodynamic occupation
probabilities, which determine the complete destruction of
an atom with increasing pressure. They should not be
confused with the optical occupation probabilities, which
determine  dissolution of spectral lines because of
the Stark shifts in plasma microfields (see
\citealp{P96b} for discussion and references).

Equation \req{Saha-m} was applied to modeling partially
ionized atmospheres of NSs, composed of iron,
oxygen, and neon \citep{RRM97,MoriHailey06,MoriHo}.
The effects related to the finite nuclear masses
(Appendix~\ref{sect:motion}) were either ignored or treated as
a small perturbation. A more accurate treatment, which
rigorously takes these effects into account, is outlined in
Appendix~\ref{sect:ioneqmotion}.

\section{Formation of spectra in strongly magnetized atmospheres}
\label{sect:atm}

\subsection{Radiative transfer in normal modes}
\label{sect:RTEmag}

Propagation of electromagnetic waves in magnetized plasmas
was studied in many works, the book by  
\citet{Ginzburg} being the most complete of them. At
radiation frequency $\omega$ much larger than the electron
plasma frequency
$\ompe=\left({4\pi e^2 n_\mathrm{e} / \mel^\ast } \right)^{1/2}$, 
where $\mel^\ast \equiv \mel \sqrt{1+\pF^2/(\mel c)^2}$,
the waves propagate in the form of two polarization modes,
extraordinary (hereafter denoted by subscript
$j=1$) and ordinary ($j=2$). They have different
polarization vectors $\bm{e}_j$ and different absorption and
scattering coefficients, which depend on the angle
$\theta_B$ between $\bm{B}$ and the wave vector $\bm{k}$. The modes interact
with each another through scattering.
\citet{GP73} formulated the radiative
transfer problem in terms of these modes. They showed that
in strongly magnetized NS atmospheres a strong Faraday depolarization occurs, 
except for narrow frequency ranges near resonances. Therefore, it is
sufficient to consider specific intensities of the two
normal modes instead of the four components of the Stokes
vector. The radiative transfer equation for these specific intensities is
\citep{KaminkerPS82}
\begin{equation}
 \cos\theta_k \frac{\dd I_{\omega,j}(\khat)}{\dd \ycol} =
    \opac_{\omega,j}(\khat) I_{\omega,j}(\khat) 
-
 \opac_{\omega,j}^\mathrm{a}(\khat)
     \frac{\mathcal{B}_{\omega,T}}{2}
-
      \sum_{j'=1}^2 \int_{(4\pi)} 
        \opac_{\omega,j'j}^\mathrm{s}(\khat',\khat)
         I_{\omega,j'}(\khat') \,\dd\khat',
\label{RTEmag}
\end{equation}
where $\khat=\bm{k}/|k|$ is the unit vector along the wave vector $\bm{k}$,
$\ycol = \int_r^\infty (1+z_g)\,\rho(r)\dd r$ is the column
density, and the factor $(1+z_g)$ is the
relativistic scale change in the gravitational field, $z_g$
being the surface redshift (see Sect.~\ref{sect:GR}).  Here, $I_{\omega,j}$ denotes the specific intensity 
of the polarization mode $j$ per unit circular frequency (if $I_\nu$ is the specific intensity per
unit frequency, then $I_\omega=I_\nu/(2\pi)$;
see~\citealt{Zheleznyakov}),
$
\mathcal{B}_{\omega,T} = 
({\hbar\omega^3}/{4\pi^3c^2})
\left(\mathrm{e}^{\hbar\omega/ T}-1\right)^{-1}
$
is the specific intensity of nonpolarized blackbody
radiation, and
\begin{equation}
\opac_{\omega,j}(\khat) \equiv
\opac_{\omega,j}^\mathrm{a}(\khat) + \sum_{j'=1}^2
\int_{(4\pi)}
\opac_{\omega,j'j}^\mathrm{s}(\khat',\khat)\,\dd\khat',
\end{equation}
The dependence of the absorption and scattering opacities $\opac^\mathrm{a}, \opac^\mathrm{s}$ on ray directions
$(\khat,\khat')$ is affected by $\bm{B}$.
Therefore, the emission of a magnetized
atmosphere, unlike the nonmagnetic case, depends not only on
the angle  $\theta_k$ between the ray and the normal direction to
the stellar surface, $\theta_\mathrm{n}$, but also on the
angle $\theta_\mathrm{n}$ between $\bm{B}$ and the normal,
and the angle $\varphi_k$ between the projections of $\bm{B}$ and
$\bm{k}$ onto the surface.  

In the plane-parallel limit, and assuming that the magnetic
field is constant in the thin photospheric layer, the equations 
for hydrostatic and energy balance are the same as in the absence of magnetic field (see, e.g.,
\citealp{SuleimanovPouW12}):
\begin{eqnarray}&&
   \frac{\dd P}{\dd \ycol} = g - g_\mathrm{rad},
\label{barometric}
\qquad
   g_\mathrm{rad} 
\approx
    \frac{2\pi}{c} 
   \int_0^\infty \dd\omega\,\opac_\omega
    \int_0^\pi
    \cos\theta_k\, I_\omega(\khat)
       \sin\theta_k\,\dd\theta_k.
\hspace*{2em}\label{grad}
\\&&
   \int_0^\infty\dd\omega \int_{(4\pi)} I_\omega(\khat)
      \,\cos\theta_k\,\dd\khat
      = F_\mathrm{ph},
\label{enbal}
\end{eqnarray}
where
$I_\omega=\sum_{j=1}^2 I_{\omega,j}$ is the total specific
intensity.

The diffusion equation for the normal modes was derived by
\citet{KaminkerPS82}. For the plane-parallel
atmosphere it reads \citep{Zavlin09}
\begin{equation}
   \frac{\dd}{\dd\ycol}
D_{\omega,j}
         \frac{\dd}{\dd\ycol} J_{\omega,j} =
        \bar{\opac}_{\omega,j}^\mathrm{a}\,
       \left[ J_{\omega,j} -
           \frac{\mathcal{B}_{\omega,T}}{2} \right]
+
 \bar{\opac}_{\omega,12}^\mathrm{s}
        \left[J_{\omega,j} - J_{\omega,3-j} \right].
\label{diffmag}
\end{equation}
Here,
\begin{eqnarray}&&
   \bar{\opac}_{\omega,j}^\mathrm{a} = \frac{1}{4\pi}\int_{(4\pi)}
\opac_{\omega,12}^\mathrm{a}\,\dd\khat,
\qquad
\bar{\opac}_{\omega,j}^\mathrm{s} = \frac{1}{4\pi}
\int_{(4\pi)}\dd\khat'
\int_{(4\pi)} \dd\khat\,\,
\opac_{\omega,12}^\mathrm{s}(\khat',\khat) \, ,
\\&&
   J_{\omega,j} = \frac{1}{4\pi}
    \int_{(4\pi)}
    I_{\omega,j}(\khat)\,\dd\khat,
\qquad
   D_{\omega,j} = \frac{1}{3\opac_{\omega,j}^{\mathrm{eff}}}=
\frac{\cos^2\theta_\mathrm{n}}{3\opac_{\omega,j}^\|} +
\frac{\sin^2\theta_\mathrm{n}}{3\opac_{\omega,j}^\perp},
\\&&
\left\{
   \begin{array}{c}
 (\opac_j^\|)^{-1}
\\
   (\opac_j^{\perp})^{-1\rule{0pt}{2ex}}
   \end{array}
  \right\}
 = \frac34 \int_0^\pi
\left\{
   \begin{array}{c}
2\cos^2\theta_B \\
\sin^2\theta_B
\end{array}
  \right\}
\frac{\sin\theta_B\,\mathrm{d}\theta_B}{\opac_j(\theta_B)}\,.
\label{kappa-eff}
\end{eqnarray}
The effective opacity for nonpolarized
radiation is
$
\opac^{\mathrm{eff}}={2}/(3D_{\omega,1}+3D_{\omega,2}).
$
The diffusion approximation (\ref{diffmag}) serves as a
starting point in an iterative method \linebreak\citep{SZ95}, which
allows one to solve the system (\ref{RTEmag})
more accurately.

\subsection{Plasma polarizability}

In Cartesian coordinates with the $z$-axis along
$\bm{B}$, the plasma dielectric tensor is
\citep{Ginzburg}
\begin{equation}
  \bm{\varepsilon} = \mathbf{I} + 4\pi\chi
          = 
 \left( \begin{array}{ccc}
 \varepsilon_\perp & \mathrm{i} \varepsilon_\wedge & 0 \\
 -\mathrm{i}\varepsilon_\wedge & \varepsilon_\perp & 0 \\
 0 & 0 & \varepsilon_\| 
 \end{array} \right),
\label{eps-p}
\end{equation}
where $\mathbf{I}$ is the unit tensor, 
$\bm{\chi}=\ChiH+\mathrm{i}\ChiA$ is the complex polarizability
tensor of plasma, $\ChiH$ and $\ChiA$ are its Hermitian and
anti-Hermitian parts, respectively. 
This tensor becomes diagonal in the cyclic (or rotating)
coordinates with unit vectors
$\hat{\mathbf{e}}_{\pm1}=(\hat{\mathbf{e}}_{x} \pm
\mathrm{i}\hat{\mathbf{e}}_{y})/\sqrt{2}$,
$\hat{\mathbf{e}}_{0}=\hat{\mathbf{e}}_{z}$.
Under the assumption
that the electrons and ions lose their regular velocity,
acquired in an electromagnetic wave, by collisions
with an effective frequency $\nu_\mathrm{eff}$
independent of the velocities, then the cyclic
components of the polarizability tensor are  (\citealp{Ginzburg},
Sect.~10)
\begin{equation}
   \chi_\alpha = -\frac{1}{4\pi}\,\frac{\ompe^2}{
    (\omega + \alpha \omc)\,(\omega - \alpha \omci)
          +\mathrm{i}\omega\nu_\mathrm{eff}}\,
\label{chi-elem}
\end{equation}
$(\alpha=0,\pm1)$. A more rigorous kinetic theory leads to
results which cannot be described by \req{chi-elem} with the
same frequency 
$\nu_\mathrm{eff}$ for the Hermitian and anti-Hermitian
components $\chiH_\alpha$ and $\chiA_\alpha$
(\citealp{Ginzburg}, Sect.~6).

The anti-Hermitian part of the
polarizability tensor determines the opacities:
\linebreak$\opac_\alpha(\omega)= 4\pi\omega\chiA_\alpha(\omega)/(\rho
c)$. Then the Kramers-Kronig relation gives \citep{BulikPavlov,KK}
\begin{eqnarray}
  \chiH_\alpha(\omega) &=&
   \frac{c\rho}{4\pi^2\omega}\, \bigg\{\!
    \int_0^\omega \!\big[\,\opac_\alpha (\omega+\omega')
     - \opac_\alpha (\omega-\omega')\,\big]
     \frac{\dd\omega'}{\omega'}
\nonumber\\&
   + & \int_{2\omega}^\infty \frac{\opac_\alpha(\omega')}{\omega'-\omega}
     \,\dd\omega'
   - \int_0^\infty \frac{\opac_{-\alpha}(\omega')}{\omega'+\omega}
     \,\dd\omega' \bigg\} .
\label{KK-mu}
\end{eqnarray}
Thus we can calculate the polarizability tensor
$\bm{\chi}$ from the opacities $\opac_\alpha(\omega)$.

\subsection{Vacuum polarization}
\label{sect:vacpol}

In certain ranges of density $\rho$ and frequency $\omega$,
normal-mode properties are dramatically affected by a
specific QED effect called vacuum polarization.
The influence of the vacuum
polarization on the NS emission was studied in detail
by \linebreak\citet{PavlovGnedin}. If the vacuum polarization is weak, then it can be linearly added to the
plasma polarization. Then the complex dielectric tensor can
be written as
$
 \bm{\varepsilon}' = \mathbf{I} + 4\pi\chi + 4\pi\chi^\mathrm{vac},
$
where 
$
 \bm{\chi}^\mathrm{vac} = (4\pi )^{-1}\,
 \mathrm{diag}(\bar{a}, \bar{a}, \bar{a}+\bar{q} )
$
is the vacuum polarizability tensor, and diag(\ldots)
denotes the diagonal matrix. The magnetic susceptibility of
vacuum is determined by expression 
$
 \bm{\mu}^{-1} = \mathbf{I} + \mathrm{diag}(\bar{a}, \bar{a},
 \bar{a}+\bar{m}).
$
\citet{Adler} obtained the vacuum polarizability 
coefficients $\bar{a}$, $\bar{q}$, and $\bar{m}$ that enter
these equations in an explicit form
at $b\ll1$, \citet{HH97} expressed them in
terms of special functions in the limits of $b\ll 1$ and
$b\gg1$. \citet{KohriYamada} presented their
numerical calculations. Finally, \citet{KK}
described them by simple analytic
expressions.
\begin{eqnarray}&&
 \bar{a}= - \frac{2\alphaf}{9\pi} \ln\bigg(
 1 + \frac{b^2}{5}\,\frac{
  1+0.25487\,b^{3/4}
  }{
  1+0.75\,b^{5/4}}\bigg),
\quad
 \bar{q} = \frac{7\alphaf}{45\pi}\,b^2\,\frac{
 1 + 1.2\,b
 }{
 1 + 1.33\,b + 0.56\,b^2
 },
\nonumber\\&&
 \bar{m} = - \frac{\alphaf}{3\pi} \, \frac{ b^2 }{
 3.75 + 2.7\,b^{5/4} + b^2 }.
\end{eqnarray}
The coefficients $\bar{a}$, $\bar{q}$, and $\bar{m}$ are not small
at $B\gtrsim 10^{16}$~G. In this case, the vacuum refraction
coefficients substantially differ from unity, and
the vacuum that surrounds a NS acts as a
birefringent lens, which distorts and additionally polarizes
thermal radiation \citep{HeylShaviv02,vanAdelsbergPerna09}. At
weaker $B$, the vacuum polarization results in
a resonance, which manifests in a mutual conversion of normal
modes, which will be considered in
Sect.~\ref{sect:atmodels}.

\subsection{Polarization vectors of the normal modes}

\citet{Shafranov} obtained the polarization vectors
$\bm{e}_j$ for fully ionized plasmas.
\citet{HoLai03} presented their convenient expressions in
terms of the coefficients $\varepsilon_\perp$,
$\varepsilon_\|$, $\varepsilon_\wedge$, $\bar{a}$,
$\bar{q}$, and $\bar{m}$, including the contributions of
electrons, ions, and vacuum polarization.
In the Cartesian coordinate system ($xyz$) with the $z$-axis
along the wave vector $\bm{k}$ and with $\bm{B}$ in the plane 
$x$--$z$, one has
\begin{equation}
\bm{e}_j=
\left(\begin{array}{c}
  e_{j,x} \\ e_{j,y} \\ e_{j,z}
  \end{array}\right)=\frac{1}{\sqrt{1+K_j^2+K_{z,j}^2}} \,
\left(\begin{array}{c}
 \mathrm{i} K_j \\ 1 \\ \mathrm{i} K_{z,j}
  \end{array}\right) ,
\label{eq:e}
\end{equation}
where
\begin{eqnarray}
&&\hspace*{-2em}
  K_j = \beta \left\{
   1 + (-1)^j \left[ 1 + \frac{1}{\beta^2} 
   + \frac{\bar{m}}{1+\bar{a}} \frac{\sin^2\theta_B}{\beta^2}\right]^{1/2}
   \right\},
\label{eq:K}
\hspace*{2em}\\&&
  K_{z,j} = - \frac{ 
   (\varepsilon_\perp' - \varepsilon_\|') K_j \cos\theta_B + \varepsilon_\wedge
   }{
   \varepsilon_\perp' \sin^2\theta_B + \varepsilon_\|' \cos^2\theta_B } 
   \, \sin\theta_B,
\label{eq:Kz}
\hspace*{2em}\\&&\hspace*{-2em}
 \beta = \frac{\varepsilon_\|' - \varepsilon_\perp' + \varepsilon_\wedge^2/\varepsilon_\perp' + \varepsilon_\|'
 \,\bar{m}/(1+\bar{a})
   }{
   2 \, \varepsilon_\wedge }
   \,\, \frac{ \varepsilon_\perp'}{\varepsilon_\|'}
   \,\,\frac{\sin^2\theta_B}{\cos\theta_B},
\hspace*{2em}
\end{eqnarray}
$\varepsilon_\perp' = \varepsilon_\perp + \bar{a}$, 
and
$\varepsilon_\|' = \varepsilon_\| + \bar{a} +
\bar{q}$.

\subsection{Opacities}
\label{sect:opac}

In the approximation of isotropic scattering, at a given
frequency $\omega$, the opacities can be presented in the
form \citep[e.g.,][]{KaminkerPS82}
\begin{equation}
   \opac_j^\mathrm{a} = \sum_{\alpha=-1}^1
     |e_{j,\alpha}(\theta_B)|^2 \,
        \frac{\sigma_\alpha^\mathrm{a}}{\mion},
\quad
 \opac_{jj'}^\mathrm{s} \!=\!\!
     {\frac34}
\!\!
  \sum_{\alpha=-1}^1 \!\!
     |e_{j,\alpha}(\theta_B)|^2 \,
     \frac{\sigma_\alpha^\mathrm{s}}{\mion}\int_0^\pi \!\!\!
       |e_{j',\alpha}(\theta_B')|^2\sin\theta_B'\,\mathrm{d}\theta_B',
\label{opac}
\end{equation}
where $\sigma_\alpha^\mathrm{a,s}$ are the absorption and scattering cross sections for the three basic polarizations. 
These include contributions of photon interaction with free
electrons or ions (free-free transitions) as well as  with
bound states of atoms and ions (bound-bound and bound-free
transitions). 

\begin{figure}[t]
\includegraphics[width=.48\linewidth]{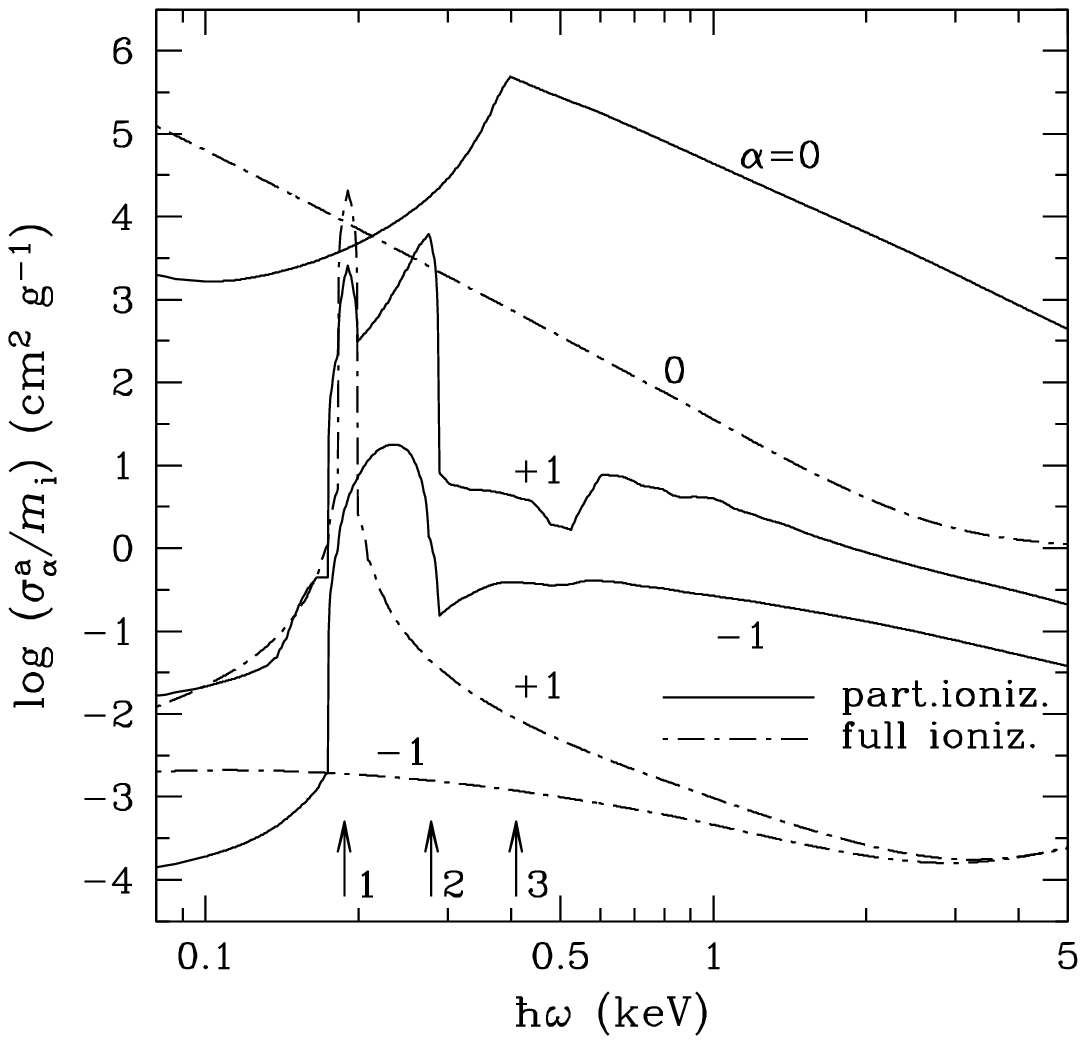}
\hspace{2ex}
\includegraphics[width=.48\linewidth]{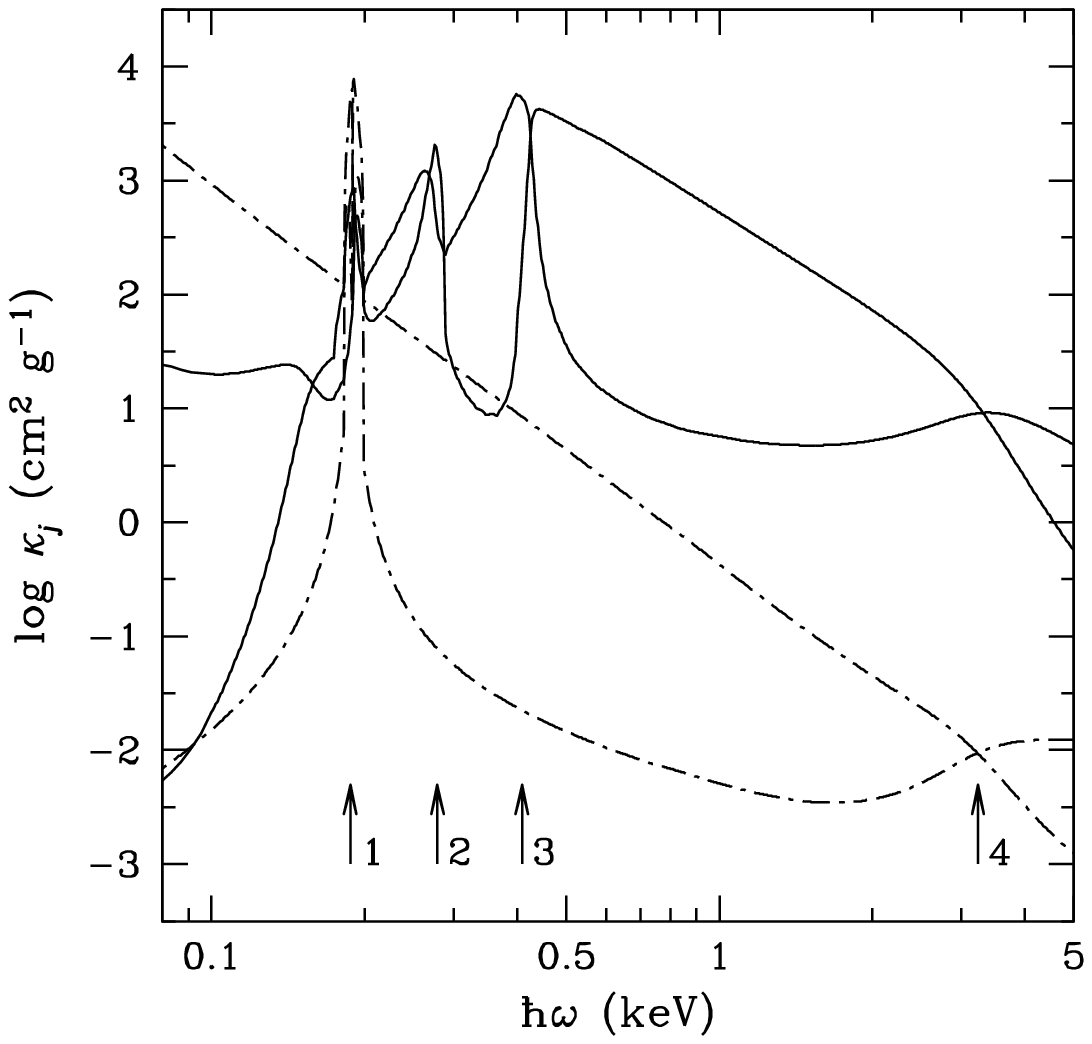}
\caption{
Logarithm of spectral opacities of a H plasma at
$B=3\times10^{13}$~G, $T=3.16\times10^5$~K, $\rho=1$
\gcc. Solid curves: partially ionized plasma model; 
dot-dashed curves: fully-ionized plasma model. 
\emph{Left panel}: basic opacities for $\alpha=0$, $\pm1$. 
\emph{Right panel}: opacities for two normal modes $j=1,2$
propagating at the angle
$\theta_B=10^\circ$ to the field lines; the
lower
(upper)
curve of each type corresponds to the extraordinary
(ordinary) wave. The arrows indicate the features at 
resonant frequencies: 1-- the ion cyclotron resonance
$\omega=\omci$; 2 -- energy threshold for a transition
between the lowest two levels
$\hbar\omega=|E_{0,0}^{(0)}-E_{1,0}^{(0)}|$;  3 -- the
ground-state binding energy $\hbar\omega=|E_{0,0}^{(0)}|$; 4
(at the right panel) -- the vacuum resonance.
\label{fig:ang13a4}}
\end{figure}

Figure~\ref{fig:ang13a4} shows basic opacities
($\sigma_\alpha^\mathrm{a}/\mion$ in \req{opac}) at $B=3\times10^{13}$~G, $\rho=1$ \gcc{} and
$T=3.16\times10^5$~K (the left panel) and corresponding
normal-mode absorption opacities $\opac_j^\mathrm{a}$ for
$\theta_B=10^\circ$. One can clearly distinguish the
features reflecting the peaks at the ion cyclotron frequency
and the resonant atomic frequencies, and the line crossings
related to the behavior of the plasma polarizability as
function of frequency. For comparison, we show also
opacities for the fully ionized plasma model under the same
conditions. They miss the features related to the atomic
resonances, and their values is underestimated by orders of
magnitude in a wide frequency range. In the remaining of this subsection we 
extend our discussion about the form of the different contributions to the opacity.

\subsubsection{Scattering}

Scattering cross-sections in NS atmospheres are well known
(\citealp{Ventura79}; \linebreak\citealp{KaminkerPS82,Mesz}).
For $\alpha=-1$, the photon-electron scattering has a
resonance at $\omc$. Outside a
narrow (about the Doppler width) frequency interval around
$\omc$, the cross sections for the basic polarizations
$\alpha=0,\pm1$ are
\begin{equation}
    \sigma_\alpha^\mathrm{s,e} =
          \frac{\omega^2}{(\omega+\alpha\omc)^2
             +\nu_{\mathrm{e},\alpha}^2}\, \sigma_\mathrm{T},
\label{sigma-se}
\end{equation}
where $\sigma_\mathrm{T}=({8\pi}/{3})({e^2}/{\mel
        c^2})^2$ is the nonmagnetic Thomson cross
section,
and
$\nu_{\mathrm{e},\alpha}$ are effective damping factors
(see below).

The photon-ion scattering cross section looks
analogously,
\begin{equation}
    \sigma_\alpha^\mathrm{s,i} =
      \left( \frac{\mel}{\mion}\right)^2
          \frac{\omega^2\,Z^4}{(\omega-\alpha\omci)^2
             +\nu_{\mathrm{i},\alpha}^2}\, \sigma_\mathrm{T}.
\label{sigma-sp}
\end{equation}
Unlike the nonmagnetic case, in superstrong fields one cannot
neglect the scattering on ions, because of the
resonance at $\omci$.
In each case, the damping factor $\nu_{\mathrm{e},\alpha}$
or $\nu_{\mathrm{i},\alpha}$
is equal to half of the total rate of spontaneous and collisional decay of the 
state with energy $\hbar\omega$ (see discussion in \citealp{PL07}). 
The  spontaneous decay rates are
\begin{equation}
    2\nu_{\mathrm{e}}^\mathrm{s} =
        \frac43\,\frac{e^2}{\mel c^3}\,\omega^2,
\quad
    2\nu_{\mathrm{i}}^\mathrm{s} =
        \frac43\,\frac{(Ze)^2}{\mion c^3}\,\omega^2.
\end{equation}
As shown by \citet{PC03} for a proton-electron plasma,
the damping factors including the scattering and 
free-free processes can be approximately written as
\begin{equation}
   \nu_{\mathrm{e},\alpha} = \nu_{\mathrm{e}}^\mathrm{s}
        + \nu_{\alpha}^{\mathrm{ff}}(\omc),
\quad
  \nu_{\mathrm{e},\alpha} = \nu_{\mathrm{e}}^\mathrm{s}
        + (\mel/\mion)\nu_{\alpha}^{\mathrm{ff}}(\omci),
\end{equation}
where $\nu_{\alpha}^{\mathrm{ff}}(\omega)$ is the effective
free-free frequency given by \req{nu-ff} below.

\subsubsection{Absorption}
\label{sect:absorption}

Without magnetic field, absorption or emission of a photon by
a free electron is impossible without involvement of another
particle, which would accept the difference between the
values of the total
momentum of the electron and the photon before and after the
absorption. In
a quantizing magnetic field, a photon can be absorbed or
emitted by a free electron in a transition between Landau levels. In the
nonrelativistic or dipole approximation, such transitions occur between the
neighboring levels at the frequency $\omc$. In the
relativistic theory, the multipole expansion leads to an
appearance of cyclotron harmonics \citep{Zheleznyakov}.
Absorption cross-sections at these harmonics were derived by
\citet{PavlovSY80} in the Born approximation without
allowance for the magnetic quantization of electron motion,
and represented in a compact form by
\citet{SuleimanovPW12}. 

The quantization of electron motion leads to
the appearance of cyclotron harmonics in the nonrelativistic
theory as well. \citet{PavlovPanov} derived photon
absorption cross-sections for an electron which moves in a magnetic field and
interacts with a nonmoving point charge. This model is
applicable at $\omega \gg \omci$. In the superstrong field
of magnetars, the latter condition is unacceptable. A more
accurate treatment of absorption of a photon by the system
of a finite-mass ion and an electron yields \citep{PC03,P10}
\begin{equation}
   \sigma_\alpha^\mathrm{ff}(\omega)
   =
          \frac{4\pi e^2
          }{ 
     \mel c} \,
   \frac{\omega^2\,\nu_{\alpha}^{\mathrm{ff}}(\omega)
          }{
          (\omega+\alpha\omc)^2 (\omega-\alpha\omci)^2
             +\omega^2 \tilde\nu_\alpha^2(\omega)},
\label{sigma-fit0}
\end{equation}
where $\nu_{\alpha}^{\mathrm{ff}}$ is an effective
photoabsorption collision
frequency, and $\tilde\nu_\alpha$ is a damping factor. In
the electron-proton plasma, taking into account the
scattering and free-free absorption, we have \citep{PC03}
\begin{equation}
   \tilde\nu_\alpha =
     \left(1+\alpha{\omc}/{\omega}\right)
     \nu_{\mathrm{i},\alpha}(\omega)
     + \left(1-\alpha{\omci}/{\omega}\right)
     \nu_{\mathrm{e},\alpha}(\omega)
       + \nu_{\alpha}^{\mathrm{ff}}(\omega).
\end{equation}
We see from
(\ref{sigma-fit0}) that
$\sigma_{-1}^\mathrm{ff}$ and
$\sigma_{+1}^\mathrm{ff}$ have a resonance at the
frequencies $\omc$ and $\omci$, respectively. 
The effective free-free absorption frequency can be written
as 
\begin{equation}
      \nu_{\alpha}^{\mathrm{ff}}(\omega) =
        \frac{4}{3}\,\sqrt{\frac{2\pi}{\mel T}}\,
          \frac{n_\mathrm{e}\, e^4}{\hbar \omega}
           \Lambda_{\alpha}^{\mathrm{ff}}(\omega),
\label{nu-ff}
\end{equation}
where
$\Lambda_{\alpha}^{\mathrm{ff}}(\omega)$
is a dimensionless Coulomb logarithm 
($\Lambda_{\alpha}^{\mathrm{ff}}=(\pi/\sqrt3)g_{\alpha}^{\mathrm{ff}}$,
where $g_{\alpha}^{\mathrm{ff}}$ is a
Gaunt factor).
Without the magnetic field, $\Lambda_{\alpha}^{\mathrm{ff}}$ is a smooth
function of $\omega$. In a quantizing magnetic field, however,
it has peaks at the
multiples of both the electron and ion cyclotron frequencies for
all polarizations
$\alpha$. However, these two types of peaks are different.
Unlike the electron
cyclotron harmonics, the ion cyclotron
harmonics are so weak that they can be safely neglected in
the NS atmospheres \citep[see][]{P10}.

\subsubsection{Bound-bound and bound-free transitions.}

The calculation of the cross section of photons with bound
states of atoms and ions is very complex. It implies
averaging of the cross sections of photon and atom
absorption over all values of $K_\perp$. Since the
distribution over $K_\perp$ is continuous for the atoms and 
discrete for the ions, such averaging for atoms reduces to
an integration over $K_\perp$, analogous to \req{Z-int},
whereas for ions it implies summation with an appropriate
statistical weight. To date, such calculation has been
realized for atoms of hydrogen \citep{PC03,PC04} and helium
\citep{MoriHeyl}.  In the Appendix, we briefly discuss
different issues related to bound states and their 
interaction with photons with the account of atomic thermal
motion in strong magnetic fields.

\subsection{Spectra of magnetic atmospheres}
\label{sect:atmodels}

\citet{Shibanov-ea92} calculated spectra
from strongly magnetized NS atmospheres,
which was assumed to be fully ionized. Later they created a database
of model spectra \citep{Pavlov95} and included it in the
\textit{XSPEC} package \citep{XSPEC}  under the name
\textsc{NSA}. They have shown that
the spectra of magnetic hydrogen and helium atmospheres are
softer than the respective nonmagnetic spectra, but harder
than the blackbody spectrum with the same temperature. In addition to the spectral
energy distribution, these authors have also studied the
polar diagram and polarization of the outgoing emission,
which proved to be quite nontrivial because of
redistribution of energy between the normal modes.  The
thermal radiation of a magnetized atmosphere is strongly
polarized, and the polarization sharply changes at the
cyclotron resonance with increasing frequency. At contrast
to the isotropic blackbody radiation, radiation of a
magnetic atmosphere consists of a narrow ($<5^\circ$) pencil beam
along the magnetic field and a broad fan beam
with typical angles  $\sim20^\circ-60^\circ$
(\citealp{Zavlin-ea95}; see also \citealp{vanAdelsbergLai}). These
calculations have thus fully confirmed the early analysis by
\citet{GnedinSunyaev74}.
Later, analogous calculations were performed by other
research groups \citep{Zane-ea01,HoLai03,vanAdelsbergLai}.
They paid special attention to manifestations of the ion
cyclotron resonance in observed spectra, which was suggested by some
SGR/AXP data. 
\citet{LaiHo02} showed that vacuum polarization
leads to a conversion of the normal modes: a photon related to one mode transforms,
with certain probability, into a photon of the other mode
while crossing a surface with a certain critical density, depending of the photon energy as
\begin{equation}
  \rho = 0.00964\,(A/Z)\,
     (\hbar\omega/\mbox{keV})^2\, B_{12}^2/f_B^2 ~\gcc,
\label{conversion}
\end{equation}
where
$f_B^2= \alphaf b^2 / [15\pi(\bar{q}+\bar{m})]$, while $\bar{q}$
and $\bar{m}$ are the vacuum-polarization coefficients
(Sect.~\ref{sect:vacpol}); $f_B\approx1$ at $B\lesssim10^{14}$~G.
The energy $\hbar\omega$ in \req{conversion} corresponds to
the line crossing in Fig.~\ref{fig:ang13a4}, indicated by
arrow labelled `4'. It follows from \req{conversion} that 
for $B\sim10^{14}$~G this energy coincides with the ion
cyclotron energy at the density where the atmosphere is
optically thin for the extraordinary mode, but optically
thick for the ordinary mode. Under such conditions, the
mode conversion strongly suppresses the ion cyclotron
feature in the emission spectrum.

For the first computations of partially ionized atmospheres
of NSs with magnetic fields 
$B\sim10^{12}$\,--\,$10^{13}$~G  \citep{Miller92,RRM97}, the
properties of  atoms in magnetic fields were calculated by
the adiabatic Hartree-Fock method \citep{MillerNeuhauser91}.
The atomic motion was either ignored \citep{Miller92} or
treated approximately by the perturbation theory
\citep{RRM97}. Later, \citet{KK} constructed a strongly
magnetized hydrogen atmosphere model beyond the framework
of the adiabatic approximation, including partial ionization
and effects of the atomic motion. The calculated spectra
revealed a narrow absorption line at the proton cyclotron
energy and some features related to atomic transitions. As
well as in the fully ionized plasma model, the intensity has
a maximum at higher energies relative to the maximum of the
Planck function, but at lower energies relative to the
nonmagnetic H atmosphere model. Therefore, the model of a
fully-ionized atmosphere with a strong magnetic field can
yield a realistic temperature, but does not reproduce the
spectral features caused by atomic transitions.

\section{A condensed surface 
as an alternative to gaseous atmospheres}
\label{sect:cond}

\citet{Ruderman71} suggested that a strong magnetic field
can stabilize polymer chains directed along the field lines,
and that the dipole-dipole attraction of these chains may
result in a condensed phase. Later works confirmed this
conjecture, but the binding and sublimation energies turned
out to be smaller than Ruderman assumed (see
\citealp{MedinLai06b}, and references therein).

From the thermodynamics point of view, the magnetic
condensation is nothing but the plasma phase transition
caused by the strong electrostatic attraction between the
ionized plasma particles. This attraction gives a negative
contribution to pressure $P_\mathrm{ex}$, which, at low temperatures, 
is not counterbalanced until the electrons
become degenerate with increasing density. In the absence of
a magnetic field, such phase transitions were studied
theoretically since 1930s (see \citealp{PPT}, for a review). In
this case, the temperature of the outer layers of a NS
$T\gtrsim(10^5-10^6)$~K exceeds the critical
temperature $\Tc$ for the plasma phase transition. However,
we have seen in Sect.~\ref{sect:EOS} that a quantizing
magnetic field lifts electron degeneracy. As a result, $\Tc$
increases with increasing $B$, which may enable such
phase transition.

\citet{Lai01} estimated the condensed-surface density as
\begin{equation}
   \rho_\mathrm{s}\approx 561\,\eta\,A Z^{-3/5}
          B_{12}^{6/5}\mbox{~\gcc},
\label{rhos}
\end{equation}
where $\eta$ is an unknown factor of the order of unity. In
the ion-sphere model \citep{Salpeter61}, the electrons are
replaced by a uniform negative background, and the potential
energy per ion is estimated as the electrostatic
energy of the ionic interaction with the negative background
contained in the sphere of radius
$\aion=(4\pi\nion/3)^{-1/3}$. By equating
$|P_\mathrm{ex}|$ to the pressure
of degenerate electrons, one obtains \req{rhos} with
$\eta=1$. This estimate disregards the effects of ion correlations,
electron-gas polarizability, and bound
state formation. Applying different versions of the
Thomas-Fermi method to the treatment of the electron polarization, one gets
quite different results: for example, the zero-temperature
Thomas-Fermi results for Fe at
$10^{10}\mbox{~G}\leqslant B\leqslant 10^{13}$~G
\citep{Rognvaldsson-ea} can be described by
\req{rhos} with $\eta \approx 0.2 +
0.01/B_{12}^{0.56}$, and in a finite-temperature
Thomas-Fermi model \citep{Thorolfsson-ea} there is no
phase transition at all.

\citet{MedinLai06b,MedinLai07} estimated the condensation
energy by the density functional method and calculated the
equilibrium density of a saturated vapor of the atoms and
polymer chains of helium, carbon, and iron above the
respective condensed surfaces at $1\lesssim B_{12} \leqslant
10^3$. By equating this density to $\rhos$, they found $\Tc$
at several $B$ values. Unlike previous authors,
\citet{MedinLai06b,MedinLai07} have considered the electron
band structure in the condensed phase.
Their results for $\rho_\mathrm{s}$ can be
described by \req{rhos} with
$\eta=0.517+0.24/B_{12}^{1/5}\pm0.011$ for carbon and
$\eta=0.55\pm0.11$ for iron, and the critical temperature
can be evaluated as  $\Tc \sim
5\times10^4\,Z^{1/4}\,B_{12}^{3/4}$~K \citep{PC13}.

When magnetic field increases from $10^{12}$~G to
$10^{15}$~G, the cohesive energy, calculated by
\citet{MedinLai07} for the condensed surface, varies
monotonically from 0.07 keV to 5 keV for helium, from 0.05
keV to 20 keV for carbon, and from 0.6 keV to 70 keV for
iron. The power-law interpolation between these limits
roughly agrees with numerical results. The electron work function
changes in the same $B$ range from 100 eV to
$(600\pm50)$~eV. With the calculated energy values,
\citet{MedinLai07} determined the conditions of
electron and ion emission in the vacuum gap above the polar
cap of a pulsar and the conditions of gap formation, and
calculated the pulsar ``death lines''.

\subsection{Radiation from a naked neutron star}

For NSs with a liquid or solid condensed surface,
the formation of thermal spectra depends on its reflectivity, first calculated by
\citet{Itoh75} and \citet{LenzenTruemper}  under simplifying assumptions.
Detailed calculations of the reflection properties of a strongly
magnetized metallic surface were presented by
\citet{Brinkmann80} and revisited
in several more recent papers 
\citep{TurollaZD04,surfem,PerezAMP05,reflefit}.
The authors determined the normal-mode
polarization vectors in the
medium under the surface, expressed the complex
refraction coefficients as functions of the
angles $\theta_k$ and $\varphi_k$ that determine the direction
of a reflected ray, and expanded the complex electric amplitudes of the
incident, reflected, and transmitted waves over
the respective basic polarization vectors. The
coefficients of these expansions, which are found from
Maxwell boundary conditions, determine the surface reflectivity
for each incident-wave polarization, $r_{\omega,j}$.
Then the total dimensionless emissivity
equals $1-\frac12(r_{\omega,1}+r_{\omega,2})$.

The early works assumed that the ions are firmly fixed in
the crystalline lattice. More recent works
\citep{surfem,PerezAMP05,reflefit}  consider not only this
model, but also the opposite limit of free ions. The real
reflectivity of the surface lies probably between the limits
given by these two extreme models, but this problem is not yet
definitely solved.

\begin{figure}[t]
\begin{center}
\includegraphics[height=.45\linewidth]{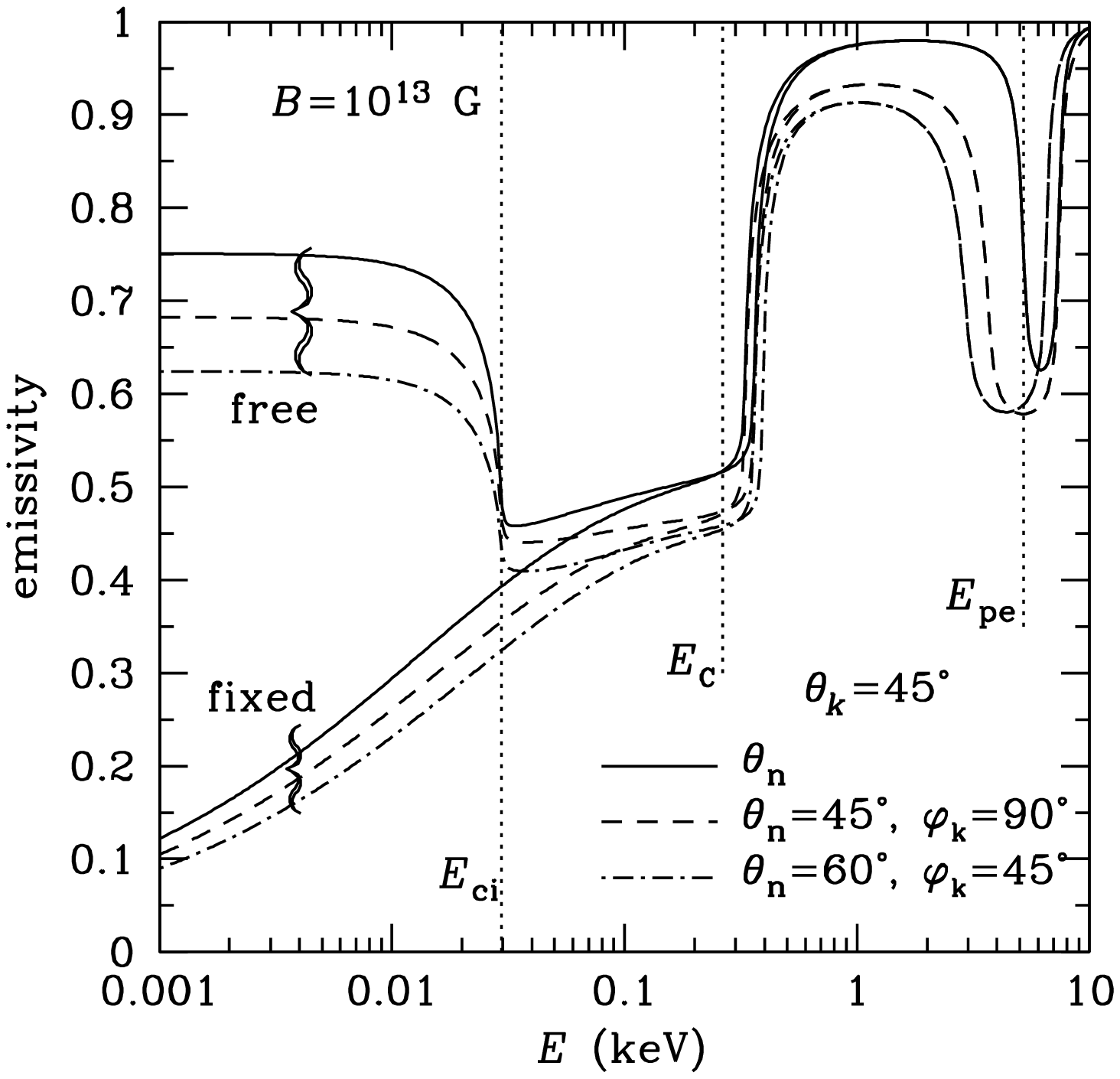}
\includegraphics[height=.45\linewidth]{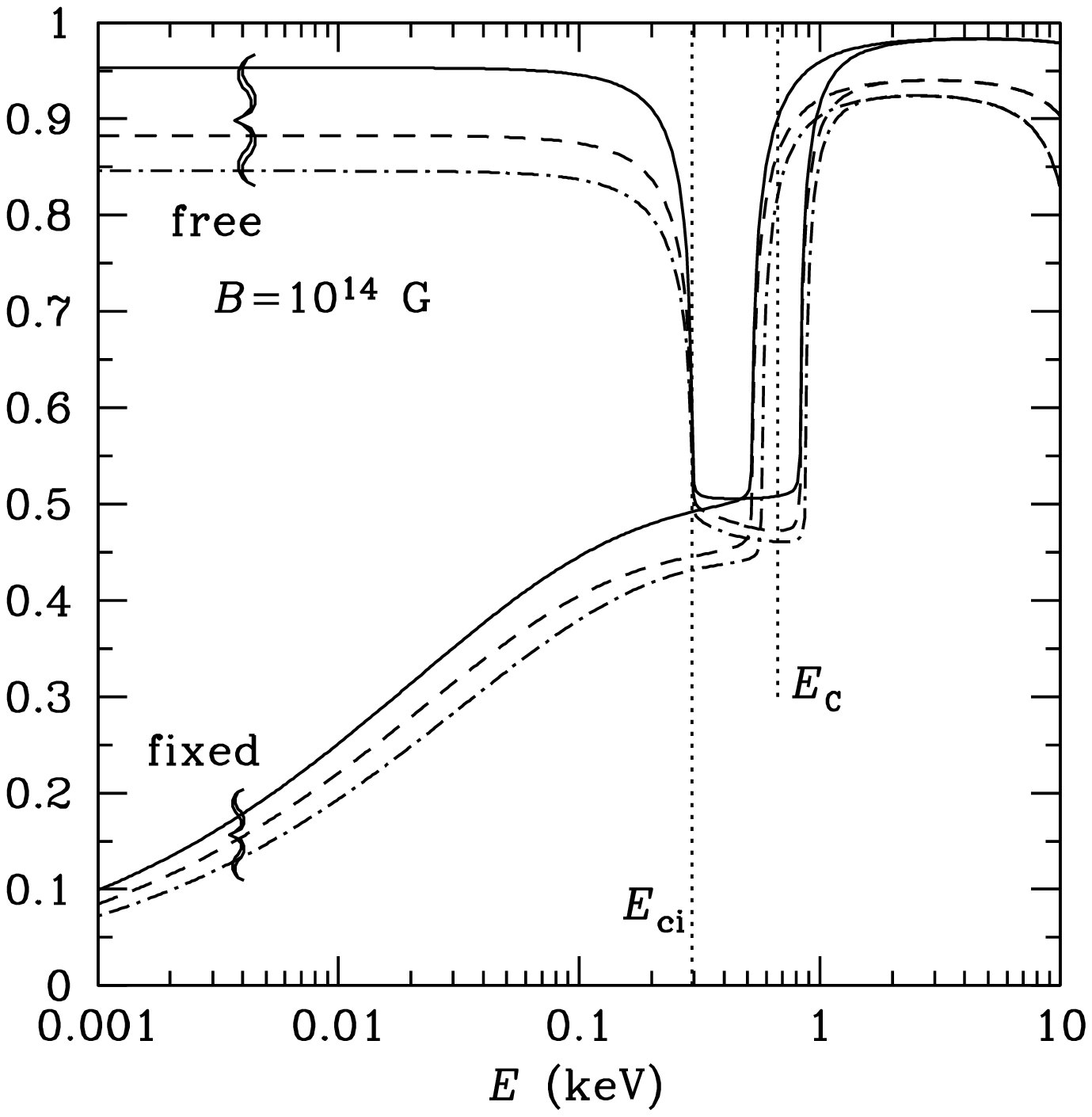}
\caption{
Dimensionless
emissivity of a condensed iron surface at $B=10^{13}$~G
(\emph{left panel}) and $10^{14}$~G (\emph{right panel}), averaged over polarizations, is shown as a
function of energy of a photon emitted at the angle
$\theta_k=45^\circ$, for different magnetic-field
inclination angles $\theta_\mathrm{n}$ and azimuthal angles
$\varphi_k$. For each parameter set, two curves are obtained
in the models of free and fixed ions. Vertical
dotted lines mark positions of the characteristic energies:
the ion cyclotron energy $E_\mathrm{ci}=\hbar\omci$, the
electron plasma energy $E_\mathrm{pe}=\hbar\ompe$, and the
hybrid energy $E_\mathrm{C}$. The groups of curves marked
``free'' and ``fixed''
correspond to the models of free and fixed ions, mentioned
in the text.
\label{fig:rb}}
\end{center}
\end{figure}

Figure \ref{fig:rb} shows examples of the emissivity, 
normalized to the blackbody intensity, as a function of the photon energy
$E=\hbar\omega$, in both the free- and fixed-ion limits,
for the wave-vector inclination angle
$\theta_\mathrm{k}=45^\circ$, $B=10^{13}$~G and $10^{14}$~G,
and different values of the magnetic-field inclination
$\theta_\mathrm{n}$ and azimuthal angles $\varphi_k$. The
characteristic energies $E_\mathrm{ci}=\hbar\omci$,
$E_\mathrm{pe}=\hbar\ompe$, and
$E_\mathrm{C}=E_\mathrm{ci}+E_\mathrm{pe}^2/\hbar\omc$ are
marked. The spectral features near these energies are
explained in \citet{surfem}. For instance, the emissivity
suppression at $E_\mathrm{ci}\lesssim E\lesssim
E_\mathrm{C}$ is due to the strong damping of one of the two
normal modes in the plasma in this energy range.
There is a resonant absorption, depending on the directions
of the incident wave and the magnetic field, near $E_\mathrm{pe}$.
The local flux density of radiation from a condensed surface
is equal to the Planck function $\mathcal{B}_{\omega,T}$
multiplied by the normalized emissivity.

In Fig.~\ref{fig:rb}, the emissivity is
averaged over polarizations. But $r_{\omega,1}\neq
r_{\omega,2}$, hence the thermal emission of a condensed
surface is polarized, the polarization depending in a
nontrivial way on the frequency $\omega$ and angles
$\theta_\mathrm{n}$, $\theta_k$, and $\varphi_k$. For
example, the degree of linear polarization can reach tens
percent near the frequencies $\omci$ and $\ompe$, which
makes promising the polarization diagnostics of NSs
with condensed surfaces. The intensity and the
polarization degree can be evaluated using analytical
approximations
for a condensed iron surface at
$B=10^{12}$\,--\,$10^{14}$~G \citep{reflefit}.

\subsection{Thin and layered atmospheres}
\label{sect:thin}

\citet{MotchZH03} suggested that some NSs can
possess a hydrogen atmosphere of a finite thickness above
the solid iron surface. If the optical depth of such
atmosphere is small for some wavelengths and large for other
ones, this should lead to a peculiar spectrum, different
from the spectra of thick atmospheres. \citet{Ho-ea07} and
\citet{SuleimanovPW09,Suleimanov-ea10} calculated such
spectra using simplified boundary conditions for the
radiative transfer equation at the inner boundary of the
atmosphere. More accurate boundary conditions
\citep{reflefit} take
into account that an extraordinary or ordinary wave, falling
from the atmosphere on the surface, gives rise to reflected waves
of both normal polarizations, whose intensities add to the
respective intensities of the waves emitted by the condensed
surface.

In general, local spectra of radiation emitted by thin
hydrogen atmospheres over a condensed surface reveal a
narrow absorption line corresponding to the proton cyclotron
resonance in the atmosphere, features related to atomic
transitions broadened by motion effects
(Appendix~\ref{sect:motion}), and a kink corresponding to the
ion cyclotron energy of the substrate ions.  Some of these
features may be absent, depending on the atmosphere
thickness and magnetic field strength. At high energies, the
spectrum is determined by the condensed-surface emission,
which is softer than the spectrum of the thick hydrogen
atmosphere.

The origin of the thin H atmospheres remains hazy.
\citet{Ho-ea07} discussed three possible scenarios. First,
it is the accretion from the interstellar medium. But its
rate should be very low, in order to accumulate the 
hydrogen mass $4\pi R^2\ycol \sim 10^{-20} M_\odot$ in
$\sim10^6$ years. Another scenario assumes diffusive nuclear
burning of a hydrogen layer accreted soon after the
formation of the NS
\citep{ChiuSalpeter64,ChangBildsten03}. But this process is
too fast at the early cooling epoch, when the star is
relatively hot, and would have rapidly consumed all the
hydrogen on the surface
\citep{ChiuSalpeter64,ChangBildsten04}. The third
possibility is a self-regulating mechanism, driven by
nuclear spallation in collisions with ultrarelativistic
particles at the regions of open field lines \citep{Jones78}.
An estimate of the penetration depth of the magnetospheric
accelerated particles indicates that this process could create
a hydrogen layer of the necessary thickness
$\ycol\sim1$ g cm$^{-2}$ \citep{Ho-ea07}.

It is natural to consider also an atmosphere having a helium
layer beneath the hydrogen layer. Indeed, all three
scenarios assume that a hydrogen-helium mixture appears
originally at the surface, and the strong gravity quickly
separates these two elements. Such ``sandwich atmosphere''
was considered by \citet{SuleimanovPW09}, where the authors
showed that its spectrum can have two or three absorption
lines in the range $E\sim(0.2$\,--\,1) keV at $B\sim10^{14}$~G.

\section{Modeling observed spectra}
\label{sect:GR}

In order to apply the local spectra models described in
Sects.~\ref{sect:atm} and \ref{sect:cond} to observations,
one has to calculate a synthetic spectrum, which would be
observed at a large distance $D$ from the star. Such
calculation should include the effects of General
Relativity, which are significant. 

The photon frequency, which equals $\omega$ in the local
inertial reference frame, undergoes a redshift
to a smaller frequency
$\omega_\infty$ in the remote observer's reference frame. 
Therefore a thermal spectrum with effective temperature
$\Teff$, measured by the remote observer,
corresponds to a lower effective temperature
\begin{equation}
   \Teff^\infty = \Teff / (1+z_g),
\label{Tinfty}
\end{equation}
where
$
 z_g \equiv \omega/\omega_\infty -1  = (1-\xg)^{-1/2} -1
$
is the redshift parameter and the compactness 
parameter $x_g=2{GM}/{c^2 R}$ of a typical NS
lies between 1/5 and 1/2. Here and hereafter the symbol
$\infty$ indicates that the quantity is measured at a
large distance from the star and can differ from its value
near the surface. 

Along with the radius $R$ that is determined by the
equatorial length $2\pi R$ in the local reference frame, one
often considers an \emph{apparent radius} for a remote
observer, 
\begin{equation}
   R_\infty = R \,(1+z_g).
\label{Rinfty}
\end{equation}
With decreasing $R$, $z_g$ increases so that the
apparent radius has a minimum, $\min R_\infty
\approx12$\,--\,14 km (\citealp{NSB1}, Chapt.~6).

The apparent
photon luminosity $L_\mathrm{ph}^\infty$ and the luminosity
in the stellar reference frame $L_\mathrm{ph}$
are determined by the Stefan-Boltzmann law
\begin{equation}
  L_\mathrm{ph}^\infty=4\pi\sSB\,R_\infty^2
          \,(\Teff^\infty)^4,
\quad
  L_\mathrm{ph} = 4\pi \sSB\,R^2 \Teff^4
\label{LSB}
\end{equation}
with $\sSB = \pi^2/(60\hbar^3 c^2)$ and $\Teff^\infty$ in
energy units.
According to (\ref{Tinfty})\,--\,(\ref{Rinfty}),
\begin{equation}
   L_\mathrm{ph}^\infty = (1-\xg)\, L_\mathrm{ph} =
   L_\mathrm{ph}/(1+z_g)^2.
\label{Linfty}
\end{equation}
In the absence of spherical symmetry, it is
convenient to define a local effective surface
temperature $\Ts$ by the relation
$
   F_\mathrm{ph}(\theta,\varphi) = \sSB\Ts^4,
$
where $F_\mathrm{ph}$ is the local radial
flux density at the surface point, determined by the polar
angle ($\theta$) and azimuth ($\varphi$) in the spherical
coordinate system. Then
\begin{equation}
   L_\mathrm{ph} = 
\int_0^{\pi}\sin\theta\,\dd\theta
\int_0^{2\pi}\dd\varphi\,R^2 F_\mathrm{ph}(\theta,\varphi)\,.
\label{Lintegral}
\end{equation}
The same relation connects the apparent luminosity
$L_\mathrm{ph}^\infty$ (\ref{Linfty}) with the apparent flux
$F_\mathrm{ph}^\infty=\sSB\,(\Ts^\infty)^4$ in the remote
system, in accord with the relation
$\Ts^\infty = \Ts/(1+z_\mathrm{g})$ analogous to
(\ref{Tinfty}).

The expressions (\ref{Tinfty}), (\ref{Rinfty}) and
(\ref{Linfty}) agree with the GR concepts of light ray bending
and time dilation near a massive body. If the angle between
the wave vector $\bm{k}$ and the normal to the surface
$\mathbf{n}$ at the emission point is $\theta_k$, then the
observer receives a photon whose wave vector makes an
angle $\theta > \theta_k$ with $\mathbf{n}$. The rigorous
theory of the influence of the light bending near a star on
its observed spectrum has been developed by
\citet{PechenickFC83} and cast in a convenient form by
\citet{Page95} and \citet{PavlovZavlin00}.
\citet{Beloborodov02} suggested the
simple approximation 
\begin{equation}
  \cos\theta_k = \xg + (1-\xg)\cos\theta,
\label{Beloborodov}
\end{equation}
which is applicable at $\xg<0.5$ with an error within a few
percent. At $\cos\theta_k < \xg$, \req{Beloborodov} gives
$\theta > \pi/2$, which allows the observer to look behind the
NS horizon.

Magnetic fields and temperatures of NSs vary from
one surface point to another. In order to reproduce the
radiation spectrum that comes to an observer, one can use
the equations derived by \citet{PoutanenGierlinski03}
(see also \linebreak\citealp{PoutanenBeloborodov06}). In the particular
case of a slowly rotating spherical star they give the
following expression for
spectral flux density related to the circular frequency
$\omega_\infty=\omega\sqrt{1-r_g/R}$ at a large distance $D$
from the star:
\begin{equation}
   F_{\omega_\infty} = 
   (1-x_g)^{3/2}\frac{R^2}{D^2}
        \int I_\omega(\bm{k};\theta,\varphi) \,
       \cos\theta_k \sin\theta
           \,\dd\theta\,\dd\varphi,
\label{Fintegral}
\end{equation}
where the integration is performed 
under the condition $\cos\theta_k>0$. The problem is
complicated, because the surface distributions of the
magnetic field and the temperature are not known in advance.
A conventional fiducial model is the
relativistic dipole \citep{GinzburgOzernoi}, while the
temperature distribution, consistent with the magnetic-field
distribution, is found from calculations of heat transport
in NS envelopes considered by \citet{Page_this_volume} in this volume.

\begin{figure}
  \includegraphics[width=.6\textwidth]{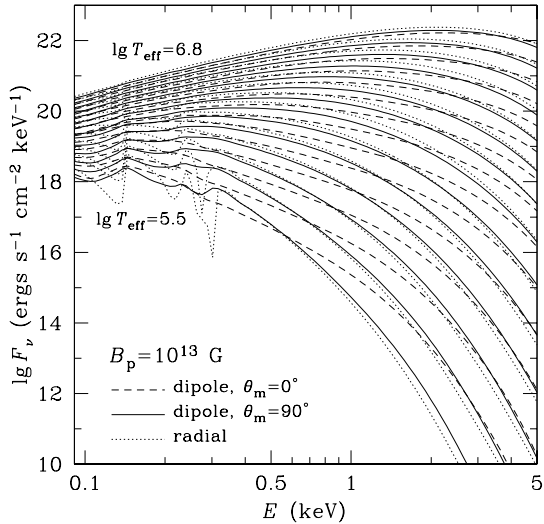}
\caption{Integral spectra of a hydrogen atmosphere of a neutron star
with $M=1.4\,M_\odot$, $R=12$ km, and with different
effective temperatures $\Teff$ ($\log\Teff$ (K) from 5.5 to
6.8 with step 0.1). The dashed and solid lines represent the
model with a dipole field of strength
$B_\mathrm{p}=10^{13}$~G at the pole and oriented along and
across the line of sight, respectively. For comparison, the
dotted curve shows the model with a constant field
$B=10^{13}$~G, normal to the surface.}
\label{fig:spectr13}
\end{figure}

For the model of partially ionized H atmospheres described
in Sect.~\ref{sect:atm}, synthetic spectra were
calculated by \citet{HoPC}. Examples are shown in
Fig.~\ref{fig:spectr13}. We see that the spectral features
are strongly smeared by the averaging over the surface, and
the spectrum depends on the magnetic axis orientation 
$\theta_\mathrm{m}$. When the star rotates, the latter
dependence leads to the phase dependence of the spectra.

Such spectra of partially ionized, strongly
magnetized NS atmospheres composed of hydrogen
have been included in the package
\textit{XSPEC} \citep{XSPEC} under the names
\textsc{NSMAX} and \textsc{NSMAXG}
(see \citealp{Ho14} and references therein).

%

\section{Theory versus observations}
\label{sect:obs}

Theoretical models of nonmagnetic atmospheres are
successfully applied to analyses of spectra of many NSs
with relatively weak magnetic fields
$B\lesssim10^9$~G. However,
the nonmagnetic models are inadequate for the strongly
magnetized NSs ($B\gg10^9$~G). The theoretical
framework for modeling the atmospheres of such stars is
described above. As argued in Sect.~\ref{sect:field-effects}, models
of strongly magnetized NS atmospheres must take
the bound species and their radiative transitions into
account. Let us consider a few examples where
models of magnetized, partially ionized atmospheres have been used to
study their thermal radiation.

\subsection{RX~J1856.5--3754}
\label{sect:1856}

In the case of RX~J1856.5$-$3754, it is surprising the absence of
absorption features and it is necessary to consider the spectrum
in the X-ray and optical ranges simultaneously. The X-ray
and optical spectra correspond to substantially
different effective temperatures if fitted separately with blackbodies. To
solve this problem, \citet{Ho-ea07} involved the
model of a thin atmosphere described in
Sect.~\ref{sect:thin}. The
measured spectrum of RX~J1856.5$-$3754 was
fitted in the entire range
from X-rays to optical within observational errorbars.
The best agreement between the theoretical and observed
spectra has been achieved at the atmosphere column density
$\ycol=1.2$ g cm$^{-2}$,
$B\sim(3$\,--\,$4)\times10^{12}$~G,
$\Teff^\infty=(4.34\pm0.03)\times10^5$~K, $z_g=0.25\pm0.05$,
and $R_\infty=17.2^{+0.5}_{-0.1}\,D_{140}$  km. Here, the
errors are given at the $1\sigma$ significance level, and
$D_{140}\equiv D/(140$ pc). Note that a fit of the observed
X-ray spectrum with the Planck function yields a 70\% higher
temperature and a 3.5 times smaller radius of the emitting
surface. Such huge difference exposes the importance of a
correct physical interpretation of an observed spectrum for
evaluation of NS parameters.

With the aid of expressions
(\ref{Tinfty})\,--\,(\ref{Rinfty}), we obtain from these
estimates $\Teff=(5.4\pm1.1)\times10^5$~K,
$R=13.8^{+0.9}_{-0.6}\,D_{140}$ km, and
$M=1.68^{+0.22}_{-0.15}\,D_{140}\,M_\odot$. \citet{Ho-ea07}
adopted the value $D=140$ pc from \citet{Kaplan-ea02}. 
Using a recent update of the distance estimate,
$D=123^{+11}_{-15}$ pc \citep{Walter-ea10}, one obtains
$R=12.1^{+1.3}_{-1.6}$ km and
$M=1.48^{+0.16}_{-0.19}\,M_\odot$. Nevertheless, the given
interpretation of the spectrum is somewhat questionable,
since it does not agree with the magnetic-field estimate
$B\approx1.5\times10^{13}$~G, obtained for this star from
timing analysis \citep{vKK08}. 

Using the same thin-atmosphere model, \citet{Ho07}
analyzed the light curve of RX~J1856.5$-$3754 and obtained
constraints on the geometry of
rotational and magnetic axes. It turned out that the light curve
can be explained if one of these angles is small
($<6^\circ$), while the other angle lies between  $20^\circ$
and $45^\circ$. In this case, the radio emission around the
magnetic poles does not cross the line of sight. As
noted by \citet{Ho07}, this may explain the
non-detection of this star as a radio pulsar
\citep{Kondratiev-ea09}.

\subsection{RX J1308.6+2127}
\label{sect:RBS1223}

\citet{Hambaryan-ea11} analyzed the X-ray spectrum
of the X-ray source RX \linebreak{}J1308.6+2127 (RBS 1223), 
which reveals a more complex structure than RX J1856.5 $-$3754.
It can be described by a wide absorption
line centered around $\hbar\omega=0.3$~keV, superposed on
the Planck spectrum, with the line parameters depending on
the stellar rotation phase. Using all 2003\,--\,2007
\textit{XMM-Newton} observations of this star, the
authors obtained a set of X-ray
spectra for different rotation phases. They tried to
interpret these spectra with different models, assuming
magnetic fields $B\sim10^{13}$\,--\,$10^{14}$~G, different
atmosphere compositions, possible presence of a condensed
surface and a finite atmosphere.

As a result, the authors managed to describe the observed
spectrum and its rotational phase dependence with the use of
the model of an iron surface covered by a partially ionized
hydrogen atmosphere with $\ycol\sim1$\,--\,10 g~cm$^{-2}$,
with mutually consistent asymmetric bipolar distributions of
the magnetic field and the temperature, with the polar
values
$B_\textrm{p1}=B_\textrm{p2}=(0.86\pm0.02)\times10^{14}$~G,
$T_\mathrm{p1}=1.22^{+0.02}_{-0.05}$ MK, and
$T_\mathrm{p2}=1.15\pm0.04$ MK. The magnetic field and
temperature proved to be rather smoothly distributed over
the surface. When compared to the theoretical model of
\citet{PerezAMP06}, such smooth distribution implies that
the crust does not contain a superstrong toroidal  magnetic
field. The effective temperature is 
$\Teff\approx0.7$ MK. The gravitational redshift is
estimated to be $z_g=0.16^{+0.03}_{-0.01}$, which converts
into $(M/M_\odot)/R_6=0.87^{+0.13}_{0.05}$ and suggests a
stiff EOS of the NS matter.

Note that the paper by \citet{Hambaryan-ea11} preceded that
by \citet{reflefit}, where the treatment of the condensed
surface and thin atmosphere was improved
(Sect.~\ref{sect:cond}). An analysis of the
spectrum of RX J1308.6+2127 
with the use of the improved results remains to be
done in the future. 

\subsection{1E~1207.4$-$5209}

The discovery of absorption lines in the spectrum of CCO 1E
1207.4$-$5209 at energies $E\sim0.7\,N$~keV ($N=1,2,\ldots$)
immediately entrained the natural assumption that they are
caused by cyclotron harmonics \citep{Bignami-ea03}. As shown
in \citet{P10}, such harmonics can be only electronic, as
the ion harmonics are unobservable. Therefore, this
interpretation implies $B\approx7\times10^{10}$~G.
\citet{MoriCH} argued that only the first and second lines
in the spectrum of 1E 1207.4$-$5209 are statistically
significant, but some authors take also the third and fourth
lines into account. This hypothesis was developed by
\citet{SuleimanovPW10,SuleimanovPW12}, who considered two
types of the electron cyclotron harmonics: the quantum
oscillations of the Coulomb logarithm and the relativistic
thermal harmonics (Sect.~\ref{sect:absorption}). An analogous
explanation of the shape of the spectrum may possibly be
applied also to CCO PSR J0821$-$4300 \citep{GotthelfHA13}.

\citet{MoriHailey06} have critically analyzed the earlier
hypotheses about the origin of the absorption lines in the
spectrum of 1E 1207.4$-$5209 and suggested their own
explanation. They analyzed and rejected such interpretations
as the lines of molecular hydrogen ions, helium ions, and
also as the cyclotron lines and their harmonics. One of the
arguments against the latter interpretation is that the
fundamental cyclotron line should have much larger depth in
the atmosphere spectrum than actually observed. Another
argument is that the cyclotron lines and harmonics have
small widths at a fixed $B$, therefore their observed width
in the integral spectrum is determined by the $B$
distribution. Thus the width of all lines should be the
same, but observations do not confirm it. These arguments of
\citet{MoriHailey06} were neglected by
\citet{SuleimanovPW10,SuleimanovPW12}. 
Both groups of authors studied the cyclotron harmonics
in spectra of fully ionized plasmas. This approach is
indeed justified for the CCOs, because the impact of bound
states on the spectra is small at $B\lesssim10^{11}$~G and
$T\gtrsim10^6$~K (Potekhin, Chabrier, and Ho, in
preparation).

As an alternative, \citet{MoriHailey06} and \citet{MoriHo}
suggested models of atmospheres composed of mid-$Z$
elements. The authors found that an oxygen atmosphere with
magnetic field $B=10^{12}$~G provides a spectrum similar to
the observed one. However, the constraint
$B<3.3\times10^{11}$~G obtained by \linebreak\citet{HalpernGotthelf}
disagrees with this model, but rather favors the
electron-cyclotron interpretation of the lines.

Unlike the cases of RX~J1856.5$-$3754 and RX J1308.6+2127 that were
considered above, there is no published results of a
detailed fitting of the observed spectrum of 1E~1207.4$-$5209
with a theoretical model. Thus all suggested explanations of
the spectrum of this object remain hypothetical.

\subsection{Rotation Powered Pulsars}
\label{sect:pulsarfit}

\subsubsection{PSR J1119$-$6127}
\label{sect:1119}

\citet{Ng-ea12} applied the partially ionized, strongly
magnetized hydrogen atmosphere model \citep{HoPC} to
interpretation of observations of pulsar J1119$-$6127, for which
the estimate based on spindown gives an atypically high
field $B=4\times10^{13}$~G. In the X-ray range, it emits
pulsed radiation, which has apparently mostly thermal
nature. At fixed $D=8.4$ kpc and $R=13$, the bolometric flux
gives an estimate of the mean effective temperature
$\Teff\approx1.1$~MK. It was difficult to explain, however,
the large pulsed fraction ($48\pm12$\%) by the thermal
emission. \citet{Ng-ea12} managed to reproduce
the X-ray light curve  of this pulsar assuming that one
of its magnetic poles is surrounded by a heated area, which
occupies 1/3 of the surface, is covered by hydrogen and
heated to $1.5$~MK, while the temperature of the opposite
polar cap is below $0.9$~MK.

\subsubsection{PSR B0943+10}
\label{sect:0943}

\citet{Storch-ea14} applied a similar analysis to
interpretation of observations of pulsar B0943+10, which
shows correlated radio and X-ray mode switches. The authors
have taken $B=2\times10^{12}$~G inferred from the pulsar
spindown, assumed $M=1.2\,M_\odot$ and $R=12$ km, and
modeled the emitting area as a hot spot covered by  a
partially ionized hydrogen atmosphere. They found that an
atmosphere with $\Teff\approx(1.4-1.5)$~MK and emission
radius $R_\mathrm{em}\approx85$~m matches the radio-quiet
X-ray spectrum, whereas previous blackbody fits gave
$T_\mathrm{bb}=3$~MK and $R_\mathrm{bb}\approx20-30$~m. The
authors showed that the large X-ray pulse fraction observed
during the radio quiet phase can be explained by including
the beaming effect of a magnetic atmosphere, while remaining
consistent with the dipole field geometry constrained by
radio observations.

\subsubsection{PSR J0357+3205}
\label{sect:0357}

A middle-aged radio-quiet gamma-ray pulsar J0357+3205 was
discovered in gamma-rays with Fermi and later in X-rays with
Chandra and XMM-Newton observatories. It produces an unusual
thermally-emitting pulsar wind nebula observed in X-rays.
\citet{Kirichenko-ea14} fitted the spectrum of this pulsar
with several different multicomponent models. In the
physically realistic case where the incomplete ionization of
the atmosphere was taken into account, they used the NSMAX
model \citep{HoPC} for the thermal spectral component and a
power-law model for the nonthermal one and fixed
$M=1.4\,M_\odot$ and $B=10^{12}$~G. They obtained an
acceptable fit ($\chi^2=1.05/244$) with a very loose
constraint on the radius, $R_\infty=8^{+12}_{-5}
(D/500\mbox{ pc})$~km.

\section{Conclusions}
\label{sec:concl}

We have considered the main features of neutron-star
atmospheres and radiating surfaces and outlined the current
state of the theory of the formation of their spectra. The
interpretation of observations enters a qualitatively new
phase, free from the assumptions of  a blackbody spectrum or
the ``canonical model'' of neutron stars. Spectral features, compatible
with absorption lines in some cases, have been discovered in thermal spectra of strongly
magnetized neutron stars. On the agenda is their detailed
theoretical description, which may provide information on the
surface composition, temperature and magnetic field
distributions. However, in order to disentangle these parameters, a number of 
problems related to the theory of magnetic atmospheres and radiating surfaces 
still have to be solved.

\begin{acknowledgements}
The work of A.P. has been partly  supported  by the RFBR 
(grant 14-02-00868)
and by the Program ``Leading Scientific Schools
of RF'' (grant NSh 294.2014.2). 
\end{acknowledgements}

\begin{appendix}

\section{The effects of finite atomic masses}
\label{sect:motion}

In this Appendix, we give a brief account of the effects of
motion of atomic nuclei in strong magnetic fields on the
quantum-mechanical characteristics of bound species and the
ionization equilibrium of partially ionized plasmas (for a
more detailed review, see \citealp{P14})

\subsection{The finite-mass effects on properties of atoms}

An atomic nucleus of finite mass, as any charged particle,
undergoes oscillations in the plane ($xy$) perpendicular
to $\bm{B}$, which are quantized in the ion Landau levels. In an atom or a molecule, these oscillations are
entangled with the electron motion. Therefore the
longitudinal projections of the orbital moments of the
electrons and the nucleus are not conserved separately.
Different atomic quantum
numbers correspond to different oscillation energies of the
atomic nucleus, multiple of its cyclotron energy
$\hbar\omci$. As a
result, the energy of every level gets an addition, which is
non-negligible if the magnetic parameter $\gamma$ is not small
compared to the nucleus-to-electron mass ratio $\mion/\mel$. 

For the
hydrogen atom and hydrogenlike ions, a conserved quantity is
$\hbar s$, which
 corresponds to the difference of
longitudinal projections of orbital moments of the atomic
nucleus and the electron, and the sum $N+s$ plays role of a
nuclear Landau number, $N$ being the electron Landau number.
For the bound states in strong magnetic fields, $N=0$,
therefore the nuclear oscillatory addition to the energy
equals $s\hbar\omci$. Thus the binding energy of a hydrogen
atom at rest is
\begin{equation}
 E_{s\nu}
    = E_{s\nu}^{(0)} - \hbar\omci s, 
\label{E0}
\end{equation}
where $E_{s\nu}^{(0)}$ is the binding energy in the
approximation of non-moving nucleus. It follows that the
values of $s$ are limited for the bound states. In
particular, all bound states have $s=0$ at $B>6\times10^{13}$~G.

The account of the finite nuclear mass is more complicated
for multielectron atoms. \linebreak\citet{AlHujajSchmelcher03} have
shown that the contribution of the nuclear motion to the
binding energy of a non-moving atom equals $\hbar\omci
S(1+\delta(\gamma))$, where ($-S$) is the total magnetic
quantum number of the atom, and $|\delta(\gamma)|\ll1$.

The astrophysical simulations assume finite
temperatures, hence thermal motion of particles. The theory
of motion of a system of point charges in a constant
magnetic field was reviewed by \citet{JHY83}. 
The canonical momentum
$\bm{P}$ is not conserved in a magnetic field. A
relevant conserved quantity is pseudomomentum 
\begin{equation}
   \bm{K}=\bm{P}+\frac{1}{2c}\,\bm{B}\times\sum_i q_i \bm{r_i}.
\end{equation}
If the system is
electrically neutral as a whole, then all Cartesian
components of $\bm{K}$ can be determined simultaneously
(i.e., their quantum-mechanical operators commute with each other). For a
charged system (an ion), one can determine $K^2$
simultaneously with either $K_x$ or $K_y$, but $K_x$ and
$K_y$ do not commute. The specific effects related to
collective motion of a system of charged particles are
especially important in NS atmospheres at
$\gamma\gg1$. In particular, so called decentered states may
become populated, where an electron is localized mostly in a
``magnetic well'' aside from the Coulomb center.

For a hydrogen atom,
$
   \bm{K} = \bm{P} - ({e}/{2c})\,\bm{B}\times \bm{r},
$
where $\bm{r}$ connects the proton and the electron. Early studies
of this particular case were done 
by \citet{GorkovDzyal,Burkova,IMS84}. Numerical
calculations of the energy spectrum of the hydrogen atom with
an accurate treatment of the effects of motion across a strong magnetic
field were performed by \citet{VDB92} and \citet{P94}.
Bound-bound radiative transitions of a moving H atom in a
plasma were studied by \citet{PP95}, and bound-free
transitions by \citet{PP97}.

\begin{figure}
\begin{center}
\includegraphics[width=.319\textwidth]{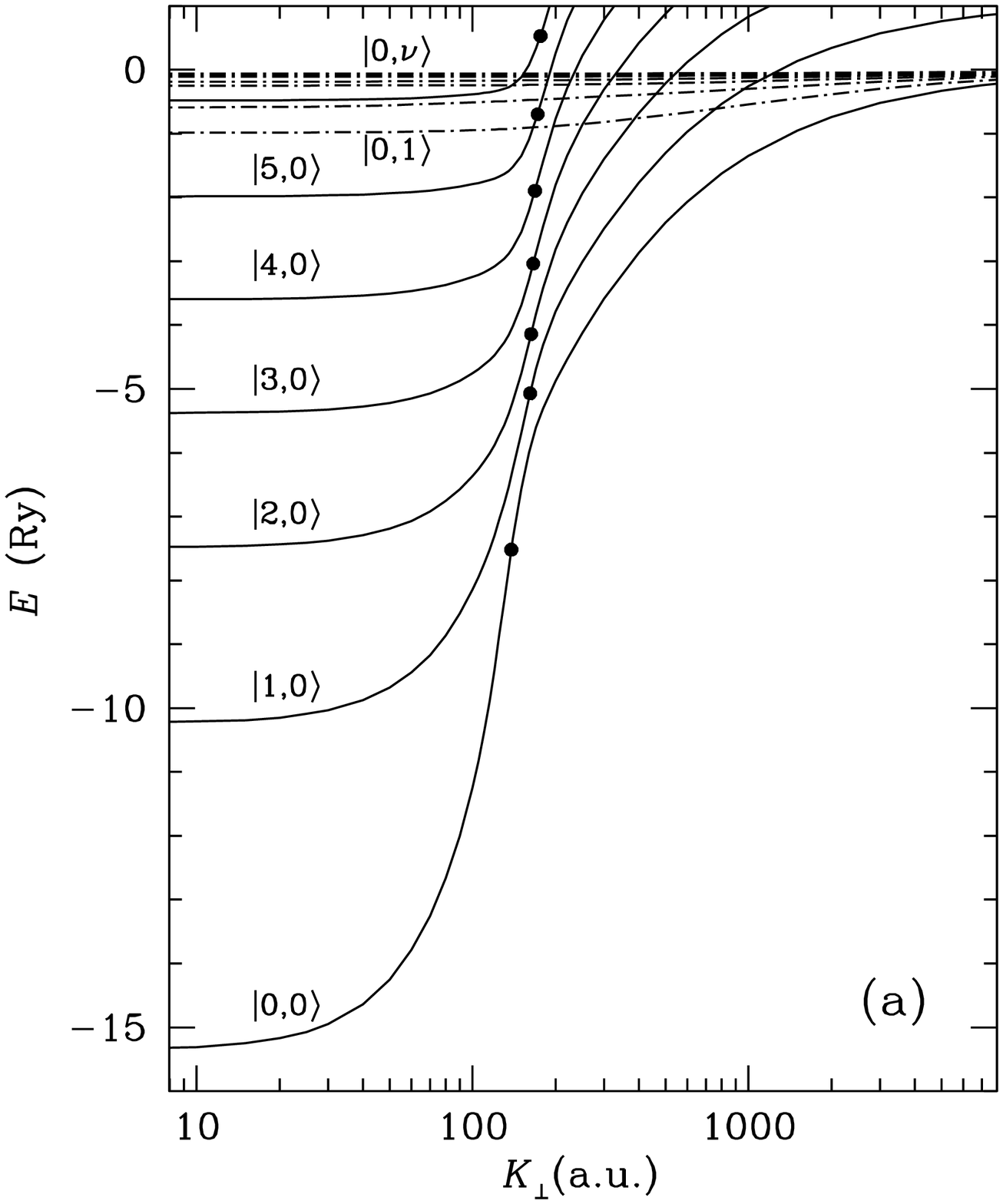}
\,
\includegraphics[width=.336\textwidth]{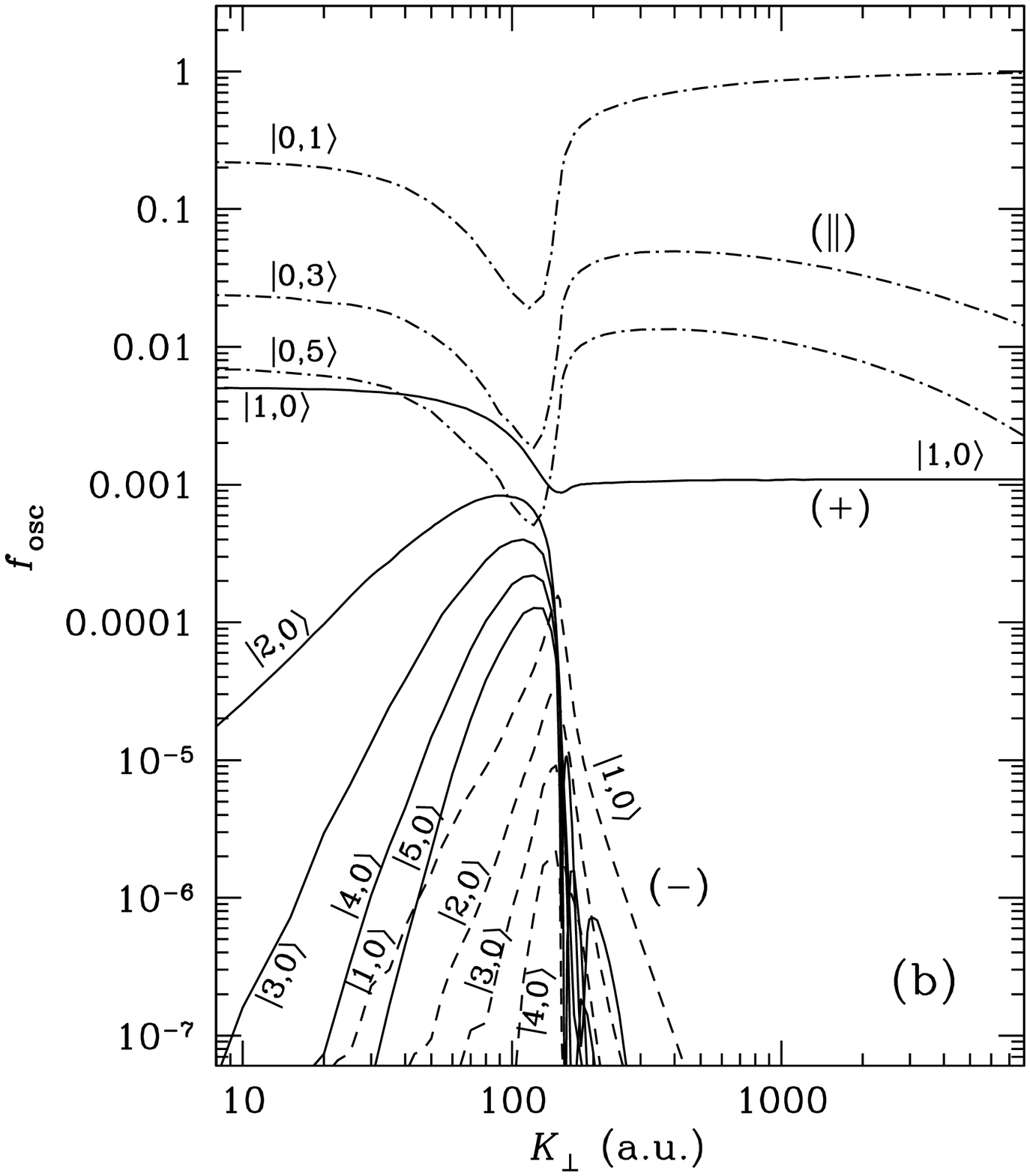}
\,
\includegraphics[width=.314\textwidth]{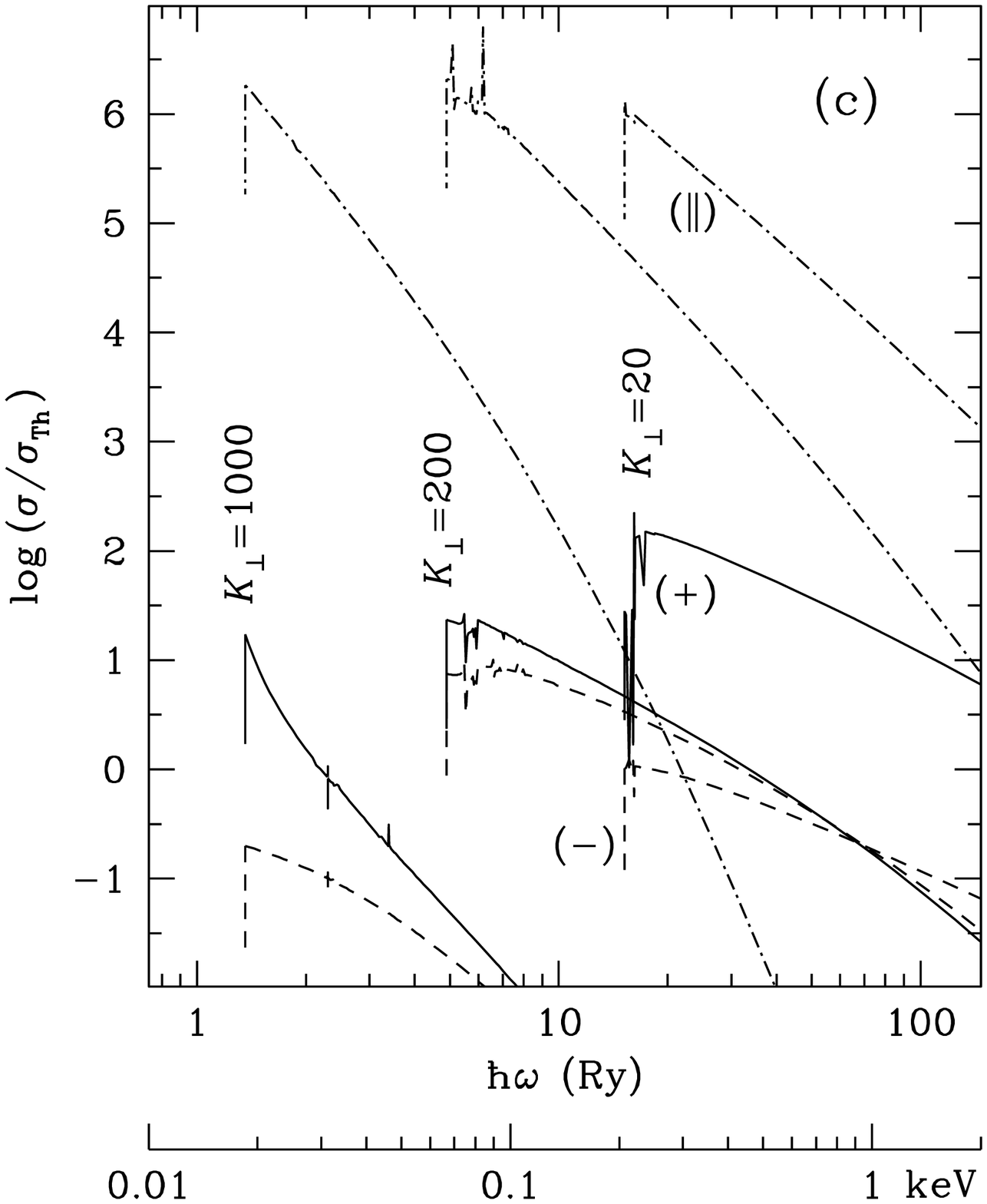}
\caption{
(a) energies, (b) oscillator strengths, and (c)
photoionization cross-sections for a hydrogen atom moving
across
magnetic field $B=2.35\times10^{12}$~G. Energies of states
$|s,0\rangle$ (solid curves) and $|0,\nu\rangle$ (dot-dashed
curves) are shown as functions of the transverse
pseudomomentum $K_\perp$ (in atomic units). The heavy dots
on the solid curves are the inflection points at
$K_\perp=\Kc$.  The  $K_\perp$-dependence of
oscillator strengths (b) is shown for transitions from the
ground state to the states $|s,0\rangle$ under influence of
radiation with polarization $\alpha=+1$ (solid curves) and
$\alpha=-1$ (dashed curves), and also for transitions into
states $|0,\nu\rangle$ for $\alpha=0$ (dot-dashed curves).
Cross sections of photoionization (c) under the influence of
radiation with $\alpha=+1$ (solid curves), $\alpha=-1$
(dashed curves), and $\alpha=0$ (dot-dashed curves) are
shown for the ground state as functions of the photon energy
in Ry (the upper x-scale) and keV (the lower x-scale) at
$K_\perp=20$~a.u.{} (the right curve),  $K_\perp=200$~a.u.{}
(the middle curve), and $K_\perp=1000$~a.u.{} (the left curve
of every type).
}
\label{fig:Hatom}
\end{center}
\end{figure}

Figure~\ref{fig:Hatom} shows the
energies, oscillator strengths, and photoionization
cross-sections of a hydrogen atom moving in a magnetic field
with $\gamma=1000$. The
reference point is taken to be the sum of the zero-point
oscillation energies of free electron and proton,
$(\hbar\omc+\hbar\omci)/2$.Therefore the negative energies in
Fig.~\ref{fig:Hatom} a correspond to bound states
($E_{s\nu}=-E>0$). At small transverse
pseudomomenta $K_\perp$, the energies of low levels in
Fig.~\ref{fig:Hatom}a exceed the binding energy of the
field-free hydrogen atom (1~Ry) by an order of magnitude.
However, the binding energy decreases with increasing
$K_\perp$, and it can become negative for the states with
$s\neq0$ due to the term $\hbar\omci s$ in \req{E0}. Such
states are metastable. In essence, they are continuum
resonances. Note that the transverse atomic velocity equals
$\partial E/\partial \bm{K}$, therefore it attains a maximum at the
inflection points ($K_\perp=\Kc$) on the curves
in Fig.~\ref{fig:Hatom}a and decreases with further increase
of  $K_\perp$, while the average
electron-proton distance continues to increase. At
$K_\perp>\Kc$ the atom
goes into the decentered state, where the electron and
proton are localized near their guiding centers, separated
by distance $r_* = (\aB^2/\hbar)K_\perp/\gamma$.

Figure~\ref{fig:Hatom}b shows oscillator strengths for the
main transitions from the ground state to
excited discrete levels. Since the
atomic wave-functions are symmetric with respect to the
$z$-inversion for the states with even $\nu$, and
antisymmetric for odd $\nu$, only the transitions that
change the parity of $\nu$ are allowed for the polarization
along the field ($\alpha=0$), and only those preserving the
parity for the orthogonal polarizations ($\alpha=\pm1$). For
the atom at rest, in the dipole approximation, due to the
conservation of the $z$-projection of the total angular
momentum of the system, absorption of a photon with
polarization $\alpha=0,\pm1$ results in the change of $s$ by
$\alpha$. This selection rule for a non-moving atom
manifests itself in vanishing oscillator strengths at
$K_\perp\to0$ for $s\neq\alpha$. In an appropriate coordinate
system \citep{Burkova,P94}, the symmetry is restored at
$K_\perp\to\infty$, therefore the transition with $s=\alpha$ is
the only one that survives also in the limit of large
pseudomomenta. But in the intermediate region of $K_\perp$,
where the transverse atomic velocity is not small, the
cylindrical symmetry is broken, so that transitions to other
levels are allowed. Thus the corresponding oscillator
strengths in Fig.~\ref{fig:Hatom}b have maxima at
$K_\perp\approx \Kc$. 
Analytical approximations for these
oscillator strengths, as well as for the dependences
of the binding energies $E_{s\nu}(K_\perp)$, are given
in \citet{P98}.

Figure~\ref{fig:Hatom}c shows photoionization cross-sections
for hydrogen in the ground state as functions of photon
energy at three values of $K_\perp$. The leftward shift of
the ionization threshold with increasing $K_\perp$
corresponds to the decrease of the binding energy that is
shown in Fig.~\ref{fig:Hatom}a, while the peaks and dips on
the curves are caused by resonances at transitions to
metastable states  $|s,\nu;K\rangle$ with positive energies
(see \citealp{PP97}, for a detailed discussion).

Quantum-mechanical calculations of the characteristics of
the He$^+$ ion that moves in a strong magnetic field are
performed by \citet{BPV97,PB05}. The basic difference from
the case of a neutral atom is that the ion
motion is restricted by the field in the transverse plane,
therefore the values of $K^2$ are quantized
\citep{JHY83}. Clearly, the similarity relations
for the ions with nonmoving nuclei (Sect.~\ref{sect:Hlike})
do not hold anymore.

Currently there is no detailed calculation of binding
energies, oscillator strengths, and photoionization
cross-sections for atoms and ions other than H and He$^+$,
arbitrarily moving in a strong magnetic field. For such
species one usually neglects the decentered states and uses
a perturbation theory with respect to $K_\perp$ (e.g.,
\citealt{MoriHailey02,MedinLP08}). This approximation can be
sufficient for simulations of relatively cool atmospheres of
moderately magnetized NSs. A condition of
applicability of the perturbation theory for an atom with
mass $m_\mathrm{a}=Am_\mathrm{u}$ requires $T/E^{(0)}\ll
m_\mathrm{a}/(\gamma \mel)\approx4A/B_{12}$ \citep{P14}. If
$B\lesssim10^{13}$~G and $T\lesssim10^6$~K, it is
satisfied for low-lying levels of carbon and heavier atoms.

\subsection{The finite-mass effects on
the ionization equilibrium and thermodynamics}
\label{sect:ioneqmotion}

Since
quantum-mechanical characteristics of an atom in a strong
magnetic field depend on the transverse pseudomomentum
$K_\perp$, the atomic distribution over $K_\perp$ cannot be
written in a closed form, and only the distribution over
longitudinal momenta $K_z$ remains Maxwellian. The first
complete account of these effects has been taken in
\citet{PCS99} for hydrogen atmospheres. Let
$p_{s\nu}(K_\perp)\,\dd^2K_\perp$ be the probability of
finding a hydrogen atom in the state $|s,\nu\rangle$ in the
element $\dd^2K_\perp$ near $\bm{K}_\perp$ in the plane of
transverse pseudomomenta. Then the number of atoms in the
element $\dd^3K$ of the pseudomomentum space equals
\begin{equation}
  \dd N(\bm{K})  = N_{s\nu}\,
        \frac{\lambda_\mathrm{a} }{2\pi\hbar}\,
         \exp\left(-\frac{K_z^2}{2m_\mathrm{a} T}\right)\, p_{s\nu}(K_\perp) \dd^3K,
\end{equation}
where $m_\mathrm{a}$ is the mass of the atom, 
$\lambda_\mathrm{a} = [{2\pi\hbar^2}/({ m_\mathrm{a} T})]^{1/2}$
is its thermal wavelength,
and $N_{s\nu}=\int\dd N_{s\nu}(\bm{K})$ is the total number
of atoms with given discrete quantum numbers. The distribution
 $N_{s\nu} p_{s\nu}(K_\perp)$ is not known in advance, but
should be calculated in a self-consistent way by
minimization of the free energy including the nonideal terms. It
is convenient to define deviations from the Maxwell
distribution with the use of generalized occupation
probabilities 
$w_{s\nu}(K_\perp)$. Then the atomic contribution
to the free energy equals
\citep{PCS99}
\begin{equation}
 F_\mathrm{id}^\mathrm{at}+F_\mathrm{int}^\mathrm{at} =
     T \sum_{s\nu} N_{s\nu}
     \int\ln\left[n_{s\nu} \lambda_\mathrm{a}^3 
        \frac{w_{s\nu}(K_\perp) }{ \exp(1) \mathcal{Z}_{s\nu}}\right]
       p_{s\nu}(K_\perp)\,\dd^2K_\perp,
\end{equation}
where
\begin{equation}
   \mathcal{Z}_{s\nu} =
\frac{ \lambda_\mathrm{a}^2 }{ (2\pi\hbar^2) } 
          \int_0^\infty w_{s\nu}(K_\perp)\mathrm{e}^{E_{s\nu}(K_\perp)/ T}
            K_\perp \dd K_\perp.
\label{Z-int}
\end{equation}
The nonideal part of the free energy that describes
atom-atom and atom-ion interactions and is responsible
for the pressure ionization has been calculated by
\citet{PCS99} with the use of the hard-sphere model. The
plasma model included also hydrogen molecules H$_2$ and
chains H$_n$, which become stable in the strong magnetic
fields. For this purpose, approximate formulae of
\citet{Lai01} have been used, which do not take full
account of the motion effects, therefore the results of
\citet{PCS99} are reliable only when the molecular fraction
is small.

This hydrogen-plasma model underlies thermodynamic
calculations of hydrogen atmospheres of NSs with
strong and superstrong magnetic fields \citep{PC03,PC04}.
\citet{MoriHeyl} applied the same approach with slight
modifications to strongly magnetized helium plasmas. One of
the modifications was the use of the plasma microfield
distribution from \citet{PGC02} for calculation of  the
$K_\perp$-dependent occupation probabilities. Mori and Heyl
considered atomic and molecular helium states of different
ionization degrees. Their treatment included 
rotovibrational molecular levels and the dependence of
binding energies on orientation of the molecular axis
relative to $\bm{B}$. The $K_\perp$-dependence of the
energy, $E(K_\perp)$, was described by an analytical fit,
based on an extrapolation of adiabatic calculations at small
$K_\perp$. The effects of motion of atomic and molecular ions
were not considered.

\end{appendix}

\end{document}